\documentclass[%
 aip,
 amsmath,amssymb,
 reprint,%
]{revtex4-1}

\usepackage{nicefrac}
\usepackage{amsthm}
\usepackage[colorlinks=true]{hyperref}

\usepackage{graphicx,color}
\usepackage{dcolumn}
\usepackage{bm}

\usepackage[utf8]{inputenc}
\usepackage[T1]{fontenc}
\usepackage{mathptmx}
\usepackage{bbm}

\newcommand{\ii}{\mathrm{i}}
\newcommand{\hc}{\mathrm{u}}
\newcommand{\dc}{\mathrm{d}}

\let \Im \relax
\DeclareMathOperator{\Im}{Im}
\let \Re \relax
\DeclareMathOperator{\Re}{Re}

\begin{document}

\preprint{AIP/123-QED}

\title{Controlling the direction of topological transport \\ in a non-Hermitian time-reversal~symmetric Floquet ladder}

\author{B. Höckendorf}
\author{A. Alvermann}%
\author{H. Fehske}
\email{fehske@physik.uni-greifswald.de}
\affiliation{%
Institut f{\"u}r Physik, Universit{\"a}t Greifswald, Felix-Hausdorff-Str. 6, 17489 Greifswald, Germany
}%


\begin{abstract}

We propose a one-dimensional Floquet ladder that possesses two distinct topological transport channels with opposite directionality. The transport channels occur due to a $\mathbb Z_2$  non-Hermitian Floquet topological phase that is protected by time-reversal symmetry. The  signatures of this phase are two pairs of Kramers degenerate Floquet quasienergy bands that are separated by an imaginary gap. We discuss how the Floquet ladder can be implemented in a photonic waveguide lattice
and show that the direction of transport in the resulting waveguide structure can be externally controlled by focusing two light beams into adjacent waveguides. The relative phase between the two light beams selects which of the two transport channels is predominantly populated, while the angles of incidence of the two light beams determine which of the transport channels is suppressed by non-Hermitian losses. 
We identify the optimal lattice parameters for the external control of transport and demonstrate the robustness of this mechanism against disorder.
\end{abstract}

\maketitle


\section{\label{sec:level1}Introduction}

The key feature of topological insulators is the existence of robust transport at the boundary which is protected by the non-trivial topology of the bulk~\cite{HasanKane2010, RevModPhys.83.1057, RevModPhys.88.035005}. In solids, non-trivial topology can emerge through the electromagnetic interaction of the Bloch electrons with external or internal gauge fields~\cite{Klitzing, Konig2007, Hsieh} and through Kramers degeneracy which is induced by fermionic time-reversal symmetry of the spin $1/2$ of the electrons~\cite{KaneMelePRL}. Naively, one would not expect that a photonic system is a suitable platform for topological insulators since photons are neutral bosons. However, both synthetic gauge fields~\cite{PhysRevLett.100.013904, Hafezi2011, RevModPhys.91.015006} and fermionic pseudo-spins~\cite{GreifRostock} can be encoded into photonic systems to realize non-trivial topology. 

In lattices of evanescently coupled waveguides~\cite{SzameitJPB}, spatially periodic modulation of the waveguides along the propagation direction has been used to achieve the encoding of gauge fields and fermionic degrees of freedom
via Floquet engineering~\cite{KitagawaPRB, Rudner, Maczewsky, Mukherjee, Rudner2020}. For example, helical waveguide arrays mimic the interaction of electrons in a solid with an external magnetic field~\cite{Rechtsman}, and Floquet protocols with pairwise coupling can faithfully reproduce the Kramers degeneracy of time-reversal~symmetric spin $1/2$ electrons~\cite{GreifRostock, HAF19}.      
Beyond the emulation of solid state phenomena, topological phases in photonic media can be deliberately subjected to non-Hermiticity~\cite{Weimann2016, Weidemann311} or non-linearity~\cite{Mukherjee856, mukherjee2020observation}, which is generally not possible in solids.

The interplay between non-Hermiticity and topology  in static~\cite{PhysRevX.8.031079, PhysRevX.9.041015, PhysRevB.99.235112, PhysRevLett.123.206404, bergholtz2020exceptional}  and Floquet~\cite{PhysRevB.98.205417, PhysRevLett.123.190403, PhysRevA.100.053608, PhysRevB.100.045423, hckendorf2019nonhermitian, Fedorova2020,PhysRevB.101.045415, PhysRevB.102.041119,PhysRevB.101.014306,H_ckendorf_2020} systems has attracted significant interest  in the last couple of years. One example of the surprising results in this field is that topological transport occurs in one-dimensional non-Hermitian chains, in direct correspondence to the boundaries of a two-dimensional topological insulator~\cite{PhysRevLett.123.206404, H_ckendorf_2020}. The topological transport in a non-Hermitian chain does not depend on fine-tuned parameters and is unaffected by moderate disorder, hence robust. It is, however, only quantized if specific conditions are satisified~\cite{hckendorf2019nonhermitian, Fedorova2020}. 
In non-Hermitian static chains, topological transport is associated with non-trivial point gaps in the complex-valued energy spectrum of the Hamiltonian~\cite{PhysRevX.8.031079}, in non-Hermitian Floquet chains with non-trivial imaginary gaps in the complex-valued quasienergy spectrum of the Floquet propagator~\cite{hckendorf2019nonhermitian, Fedorova2020}. 

In both cases, we can distinguish between non-Hermitian chains that possess a unidirectional topological transport channel and non-Hermitian chains that possess two topological transport channels with opposite directionality. Due to the robustness of topological transport, it does not matter how the unidirectional chain is initially excited: Transport always goes in the same direction. On the other hand, the preparation of the initial state in a chain with bidirectional transport determines which of the two transport channels is predominantly populated. Therefore, the direction of the transport can be externally controlled by manipulating the initial state. 

\begin{figure}
\includegraphics[width=1\linewidth]{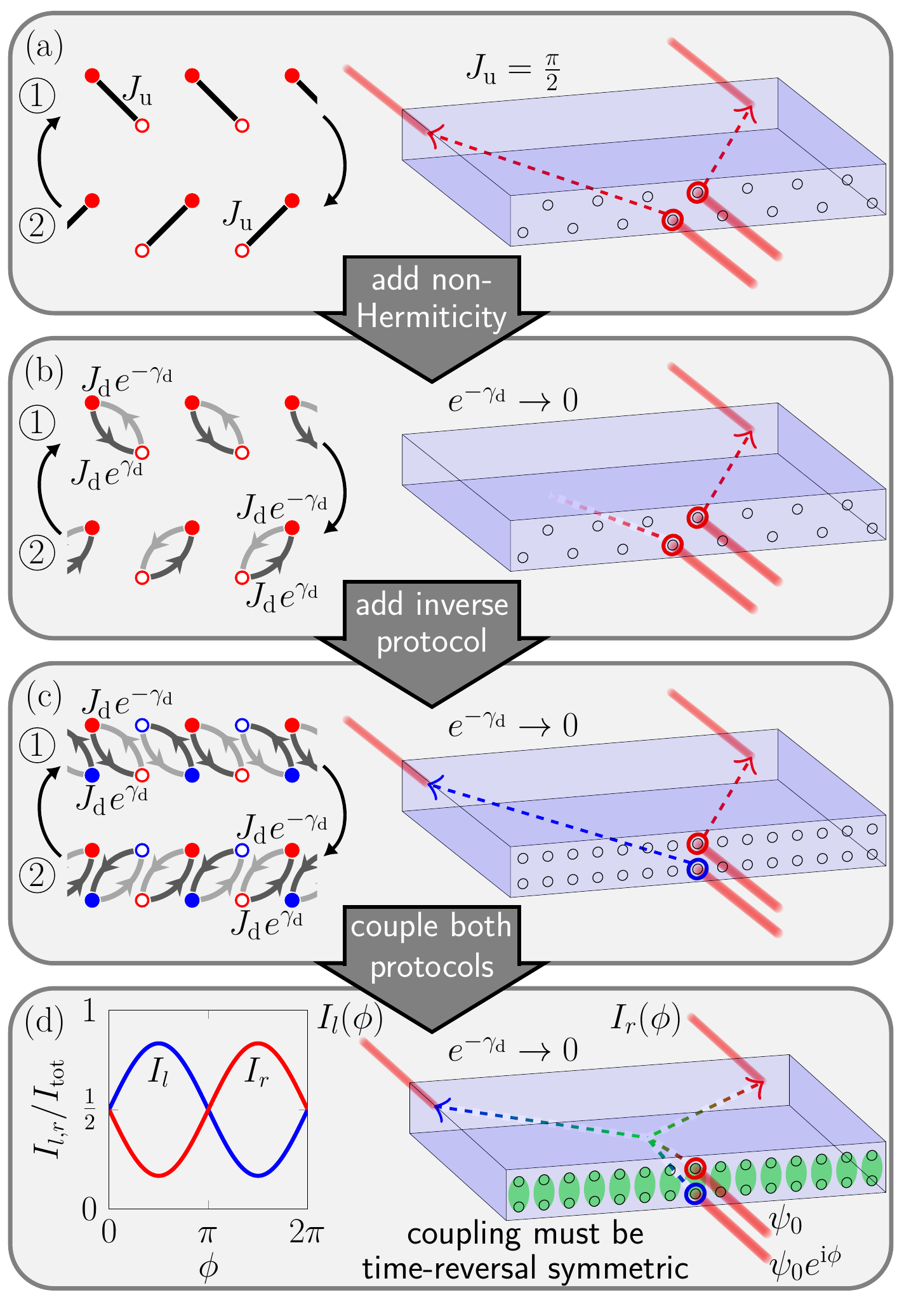}
\caption{ (a) Hermitian Floquet zigzag chain with bidirectional, but not robust, transport and its realization in a photonic waveguide lattice (blue box). Exciting the waveguides (indicated by black circles) on the top (bottom) of the front facet induces transport to the right (left). (b) By adding non-Hermitian directional coupling (light and dark gray lines), the transport becomes unidirectional and robust. (c) A second Floquet zigzag chain with inverse directionality is added. The combined Floquet ladder is time-reversal symmetric and supports two independent unidirectional transport channels with opposite directionality. (d) Time-reversal~symmetric couplings  connect the two zigzag chains along the rungs of the ladder. 
Now the direction of transport can be externally controlled by varying the relative phase $\phi$ between the two input light beams.}
\label{Fig:1}
\end{figure}

In this work, we introduce a one-dimensional non-Hermitian time-reversal~symmetric Floquet ladder that possesses two topological transport channels with opposite directionality. The Floquet ladder is constructed in such a way that (i) it can be readily implemented in photonic waveguide lattices, and (ii) the direction of transport in the resulting waveguide structure can be externally controlled with the setup sketched in Fig.~\ref{Fig:1}(d). Two waveguides that form a rung of the ladder (indicated by the green ellipses) are simultaneously excited by two light beams that have the same intensity $|\psi_0|^2$ but a variable relative phase $\phi$. 
This initial excitation populates both the left-moving and right-moving transport channel in the waveguide lattice. Depending on the specific value of $\phi$, either the left- or right-moving transport channel is predominantly populated, influencing the output intensities $I_l(\phi)$, $I_r(\phi)$ of the two channels.
By varying the relative phase, we can switch between equal transport in both directions ($I_l=I_r$), predominant transport to the left ($I_l>I_r$), or predominant transport to the right ($I_l<I_r$). The total output intensity $I_{\mathrm{tot}}=I_l(\phi)+I_r(\phi)$ remains constant during this process.

Our Floquet ladder hosts a $\mathbb Z_2$ topological phase that exclusively occurs in a non-Hermitian time-reversal~symmetric Floquet system.
Before we specify the relevant Floquet protocol, let us explain why both non-Hermiticity and time-reversal symmetry are necessary to control the direction of topological transport in a one-dimensional chain.

\section{Construction of the Floquet protocol}

The simplest one-dimensional Hermitian Floquet protocol that realizes bidirectional transport is shown in Fig.~\ref{Fig:1}(a). It is implemented on a zigzag chain with lattice sites $\textcolor{red}{\bullet}$ (filled red circles) and $\textcolor{red}{\circ}$ (open red circles). The time-periodic Hamiltonian, $H(t+T)=H(t)$, cycles through two consecutive steps, each of which couples two different adjacent sites in the zigzag chain. The uniform (u) pairwise coupling can be described by the $2\times 2$ Hamiltonian
\begin{equation}
\label{coup_h}
H_{\hc}=\begin{pmatrix} 0& J_\hc\\ J_\hc & 0\end{pmatrix}
\end{equation}
with coupling strength $J_\hc$. The associated propagator is $U_{\hc}=\exp(-\ii H_{\hc} \delta t)=\cos(J_\hc \delta t)\mathbbm{1}-\ii \sin(J_\hc \delta t) \sigma_x$, with the identity matrix $\mathbbm 1$, the Pauli matrix $\sigma_x$, and the duration of a time step $\delta t$. In the following, we set $\delta t\equiv 1$ for all time steps.  At $J_\hc=\pi/2$, we have $U_{\hc}=-\ii \sigma_x$. Consequently, a full amplitude transfer occurs between coupled sites. In this way, bidirectional transport is realized in the chain. An excitation starting on a $\textcolor{red}{\bullet}$ site ($\textcolor{red}{\circ}$ site) is transferred two sites to the right (left) in each period of the Floquet protocol. Note that this transport requires fine-tuning of the coupling strength $J_\hc$. In addition, it is not robust against disorder~\cite{hckendorf2019nonhermitian}.

Robust transport is achieved if non-Hermiticity is added to the chain by replacing the uniform pairwise coupling~\eqref{coup_h} with the directional (d) pairwise coupling
\begin{equation}
\label{coup_nh}
H_{\dc}=\begin{pmatrix} 0& J_\dc  e^{\gamma_\dc}\\ J_\dc e^{-\gamma_\dc}& 0\end{pmatrix} \;,
\end{equation}
as shown in Fig.~\ref{Fig:1}(b).
Here, the parameter $\gamma_\dc$ controls the directionality of the coupling.  
For $\gamma_\dc \ne 0$, the coupling incorporates gain for the transfer in one direction, leading to amplification of excitations, and loss for the transfer in the opposite direction, leading to suppression of excitations. Note that gain and loss in the directional  coupling need only be specified relative to the uniform loss or gain of the entire system that does not favor a specific direction. 
The associated propagator $U_{\dc}=\exp(-\ii H_{\dc})$ is now given by $U_{\dc}=\cos(J_\dc)\mathbbm{1}-\ii \sin(J_\dc) e^{\gamma_\dc} \sigma_+-\ii \sin(J_\dc)e^{-\gamma_\dc} \sigma_-$, with the Pauli matrices $\sigma_\pm= \sigma_x/2\pm\ii \sigma_y/2$. In the limit $e^{-\gamma_\dc} \to 0$, we have $U_{\dc} e^{-\gamma_\dc}=-\ii \sin(J_\dc) \sigma_+$. An excitation starting on a $\textcolor{red}{\bullet}$ site is transferred two sites to the right, irrespective of the concrete value of the coupling strength $J_\dc$. On the other hand, an excitation starting on a $\textcolor{red}{\circ}$ site is suppressed during propagation. The analogous result, with inverted roles of the two lattice sites, holds in the limit $e^{\gamma_\dc}\to 0$. To avoid redundancies, we will only consider the limit $e^{-\gamma_\dc}\to 0$. Transport is now unaffected by the addition of disorder~\cite{hckendorf2019nonhermitian}, but is also unidirectional instead of bidirectional.

Bidirectional transport is recovered if a second Floquet zigzag chain with inverse directionality is added. Here, the two chains are arranged in such a way that they form the ladder shown in Fig.~\ref{Fig:1}(c). 
On the second zigzag chain, excitations starting on $\textcolor{blue}{\bullet}$ sites (filled blue circles) get transferred to the left, while excitations starting on $\textcolor{blue}{\circ}$ sites (open blue circles) get suppressed for $e^{-\gamma_\dc} \to 0$. Since the two zigzag chains are decoupled, there is no interference between  excitations that start on adjacent filled sites $\textcolor{red}{\bullet}$, $\textcolor{blue}{\bullet}$.

To enable the external control of topological transport depicted in Fig.~\ref{Fig:1}(d), the two zigzag chains have to be coupled. Simultaneously,  the bidirectional transport has to retain its robustness. As with the boundary states in $\mathbb Z_2$ topological insulators~\cite{KaneMelePRL}, this is only possible if Kramers degeneracy is enforced upon the two counter-propagating transport channels. To achieve this, the zigzag chains will be coupled in such a way that the combined ladder satisfies the relation 
\begin{equation}
\label{TRS_t}
\theta H(t) \theta^{-1} = H^t(T-t) \; 
\end{equation} 
of fermionic time-reversal symmetry, in which the unitary time-reversal operator $\theta$ fullfills $\theta \theta^*=-1$ and maps the red and blue sites onto each other as indicated by the green ellipses in Fig.~\ref{Fig:1}(d). In this case, the red and blue sublattice can be identified as the up and down components of a fermionic pseudo-spin $1/2$. The time-reversal operator acts on this pseudo-spin as $\theta=\sigma_y$. In this pseudo-spin notation, the initial state \textcolor{blue}{$|\Psi_0\rangle$} used for the control of transport can be written as $\textcolor{blue}{\langle x|\Psi_0\rangle}=\psi_0(x) (| {\uparrow}\rangle +e^{\ii \phi} |{\downarrow}\rangle)$, where $\uparrow$ ($\downarrow$) corresponds to the upper (lower) row of the Floquet ladder. The wavefunction $\psi_0(x)$ has a non-zero value $\psi_0(x)=\psi_0$ on a single filled site and is zero everywhere else.

 In the symmetry relation~\eqref{TRS_t}, $H^t$ denotes the transpose of the Hamiltonian.
 Note that for a non-Hermitian Hamiltonian $H^t$ does not coincide with the complex-conjugate $H^*$,
 such that replacing $H^t$ with $H^*$ in Eq.~\eqref{TRS_t} gives a different type of time-reversal symmetry~\cite{PhysRevX.9.041015, PhysRevB.99.235112}. We do not consider this symmetry because it enforces unidirectional instead of bidirectional transport~\cite{PhysRevLett.123.190403}. 

Graphically, time reversal acts as follows on the Floquet ladder: (i) The order of the time steps is inverted, the directionality of each coupling line is inverted, and each coupling line is mirrored across the horizontal axis between the upper and lower row of the ladder. The Floquet protocol in Fig.~\ref{Fig:1}(c) thus satisfies the symmetry relation~\eqref{TRS_t}. Possible couplings between the red and blue sublattice additionally acquire a minus sign under time-reversal due to the constraint $\theta \theta^*=-1$.

We now couple the two zigzag chains by adding four steps to the Floquet protocol which consist of horizontal and vertical uniform pairwise couplings with coupling strength $\pm J_\hc$, see steps $2$, $3$, $4$, and $5$ in Fig.~\ref{Fig:2}(a). The four steps also satisfy Eq.~\eqref{TRS_t}. Both horizontal and vertical couplings are required to induce interference between excitations that start on adjacent red and blue sites. If, for example, we omit steps $2$ and $5$ from the protocol,  time-reversal symmetry is still satisfied but the two zigzag chains are decoupled in the limit $e^{-\gamma_\dc}\to 0$. Similarily, steps $2$ and $5$ cancel if we omit steps $3$ and $4$. 

In total, the Floquet protocol consists of six steps. Four of these steps utilize uniform pairwise couplings~\eqref{coup_h} (steps $2$, $3$, $4$, $5$)  and two steps utilize directional pairwise couplings~\eqref{coup_nh} (steps $1$, $6$). Both coupling types can be readily implemented in photonic waveguide lattices. 

\section{Photonic implementation}

\begin{figure}
\includegraphics[width=1\linewidth]{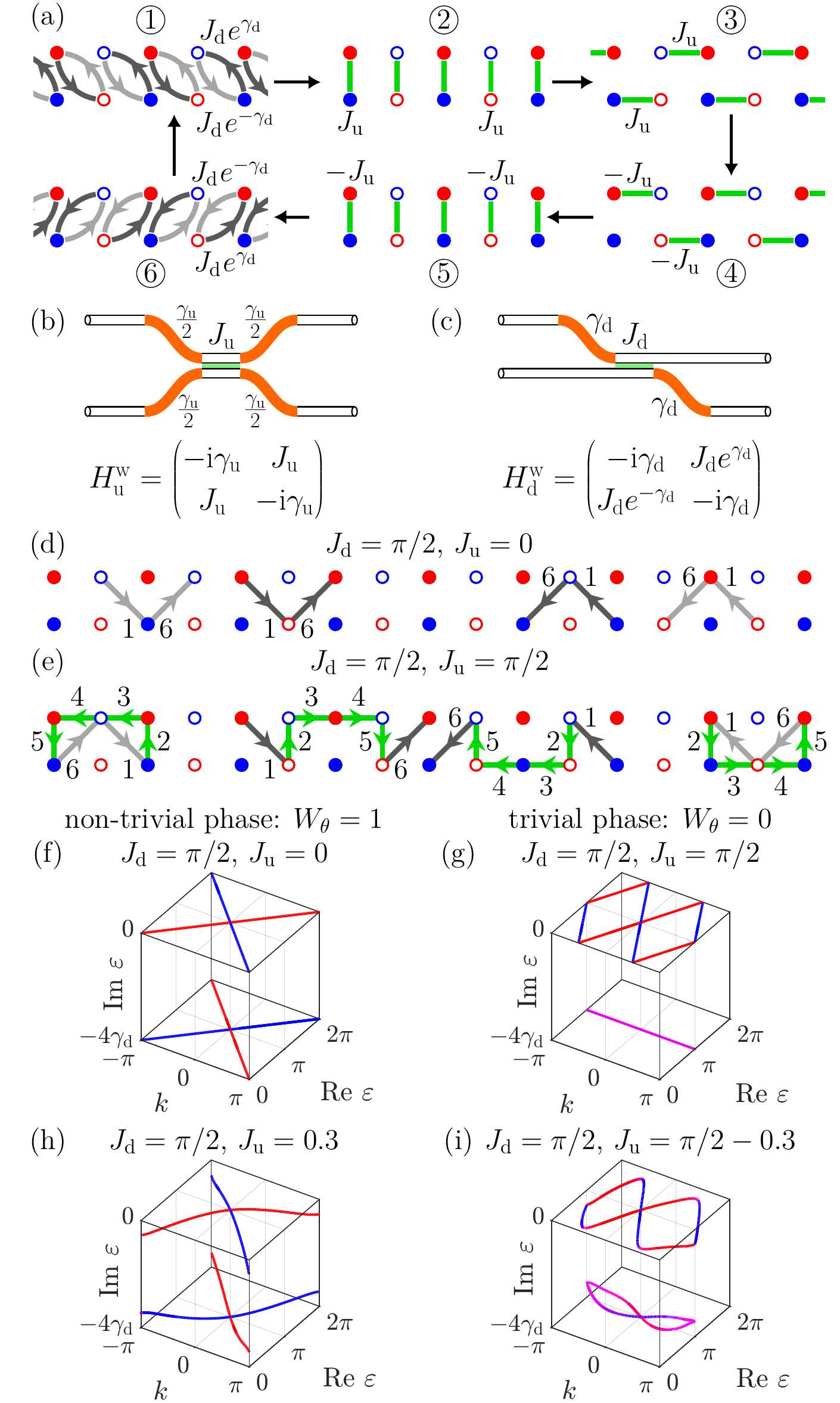}
\caption{
(a) Floquet ladder with time-reversal~symmetric coupling between the red and blue sublattice. (b) Photonic implementation of the uniform pairwise coupling in steps $2$, $3$, $4$, and $5$. (c) Photonic implementation of the directional pairwise coupling in steps $1$ and $6$. Note that the final position of the two waveguides is shifted. If necessary, the waveguides can be returned to their original position after the coupling step. (d,e)  Patterns of motion during one cycle of the Floquet protocol and (f,g) the quasienergy bands  for two specific parameter sets. The colors of the quasienergy bands indicate the spatial distribution of the corresponding eigenvectors on the red and blue sublattice. The color of the twofold degenerate flat band indicates equal distribution on red and blue lattice sites. (h,i) Testing the robustness of the windings around the quasienergy zone for $J_\hc \ne 0, \pi/2$. We use $\gamma_\dc=0.5$ in all spectra.   
}
\label{Fig:2}
\end{figure}

To implement uniform pairwise coupling, the distance between two waveguides must be locally reduced in a symmetric manner, as shown in Fig.~\ref{Fig:2}(b). In contrast to Eq.~\eqref{coup_h}, the effective Hamiltonian $H_{\hc}^{\mathrm{w}}=H_{\hc}-\mathrm{i}\gamma_\hc \mathbbm{1}$ and propagator $U_{\hc}^{\mathrm{w}}=U_{\hc} \, e^{-\gamma_\hc}$ for the uniform coupling of two waveguides (w) include additional losses $\gamma_\hc>0$ which occur due to the bending of the waveguides. 

The bending losses can be used to implement the directional coupling, as shown in Fig.~\ref{Fig:2}(c). The distance between two waveguides is again locally reduced but now in an asymmetric manner.  In comparison to Eq.~\eqref{coup_nh}, the effective Hamiltonian $H_{\dc}^{\mathrm{w}}=H_{\dc}-\mathrm{i}\gamma_\dc \mathbbm{1}$ and propagator $U_{\dc}^{\mathrm{w}}=U_{\dc}\, e^{-\gamma_\dc}$ of the waveguides are also shifted by the bending losses $\gamma_\dc>0$, 
such that there is no amplification factor $e^{\gamma \dc}$ in the propagator $U_{\dc}^{\mathrm{w}}$. Because of this, the propagator $U_{\dc}^{\mathrm{w}}=\sin(J_\dc)\sigma_+$ is finite for $e^{-\gamma_\dc}\to 0$.   In particular, directional couplings require no gain but can be realized purely with losses. This is important for passive systems such as waveguide lattices, where losses can be controlled through the curvature of the waveguides but the implementation of actual gain is significantly harder~\cite{Weimann2016,Ornigotti_2014}.  


The waveguide configuration in Fig.~\ref{Fig:2}(b) exlusively realizes positive couplings $J_\hc>0$. To implement the protocol steps with negative couplings, the periodicity $U_{\hc}^{\mathrm{w}}=(-1)^n e^{-\gamma_\dc} [\cos(J_\hc+n\pi)\mathbbm 1-\ii  \sin(J_\hc+n\pi) \sigma_x]$ of the propagator can be exploited. Any negative coupling $J_\hc<0$ can be replaced with an equivalent positive coupling $J_\hc+n\pi>0$ for large enough $n\in N$.  The additional global phase factor $(-1)^n$ is irrelevant for the light propagation in the waveguides.

\section{Topological classification}

Now, we show that the Floquet ladder hosts a $\mathbb Z_2$ non-Hermitian Floquet topological phase which is protected by time-reversal symmetry. This phase can be fully characterized by the spectrum $e^{-\ii \varepsilon(k)}$ of the Floquet-Bloch propagator $U(k)$ with the complex-valued quasienergies $\varepsilon(k)=\Re \varepsilon(k)+\ii \Im \varepsilon(k)$. Note that the real part $\Re \varepsilon(k)$ of the quasienergies is defined only up to multiples of $2\pi$. Here, we restrict it to the quasienergy zone $\Re \varepsilon \in [0,2\pi)$.

The Floquet-Bloch propagator is given by $U(k)=\prod_{i=1}^6 U^{(i)}(k)$ where $U^{(i)}(k)=\exp(-\ii \, H^{(i)}(k))$ are the Bloch propagators and $H^{(i)}(k)$ the Bloch Hamiltonians of the six steps. The Bloch Hamiltonians and Bloch propagators can be deduced from the effective Hamiltonians $H^{\mathrm{w}}_\hc$, $H^{\mathrm{w}}_\dc$ and propagators $U^w_\hc$, $U^w_\dc$ of the coupled waveguides in combination with Fig.~\ref{Fig:2}(a). Since there is no gain in the waveguide lattice, the imaginary part of the quasienergies $\Im \varepsilon(k)\leq -4\gamma_\hc$ is bounded from above.  Here, $-4\gamma_\hc$ are the uniform losses that all excitations in the Floquet ladder accumulate during one Floquet cycle due to steps $2$, $3$, $4$, and $5$. 
Since the uniform losses act equally on all excitations, they are irrelevant for our discussion. In the following, we set $\gamma_\hc\equiv 0$ and thus omit the uniform losses in the specification of quasienergies $\varepsilon(k)$ and output intensities $I_{l,r}(\phi)$.

The symmetry relation~\eqref{TRS_t} implies that the four quasienergy bands of the ladder can be divided into two Kramers pairs, each of which satisfies  
\begin{equation}
\label{symm_eigen}
\varepsilon_m^I(k)=\varepsilon_m^{II}(-k) \; ,
\end{equation}
where $m=1,2$. The quasienergies $\varepsilon_m^I$, $\varepsilon_m^{II}$ of each pair are degenerate at the invariant ($k\equiv -k$) momenta $k=0,\pi$ and the corresponding eigenvectors are biorthogonal~\cite{PhysRevX.9.041015, PhysRevB.99.235112}. The Kramers degeneracy at the invariant momenta can only be lifted by breaking time-reversal symmetry.

To illustrate the main properties of the $\mathbb Z_2$ phase, let us consider the two parameter sets $J_\dc=\pi/2, J_\hc=0$ and $J_\dc=\pi/2$, $J_\hc=\pi/2$, which realize the non-trivial and trivial phase, respectively. Figs.~\ref{Fig:2}\textcolor{blue}{(d,e)} show the patterns of motion and Figs.~\ref{Fig:2}(f,g) the quasienergy dispersions for the two parameter sets.

At $J_\dc=\pi/2$, $J_\hc=0$, any single-site excitation is transferred one unit cell to the right or left during one Floquet cycle. Correspondingly, each of the four quasienergy bands winds once around the quasienergy zone, two bands with positive chirality and two bands with negative chirality. 
The four bands form two Kramers pairs with the dispersions $\varepsilon_1^{I,II}(k)=\pm k+\pi-4\ii \gamma_\dc$ and $\varepsilon_2^{I,II}(k)=\pm k+\pi$. Due to the directional coupling, the two pairs are spectrally separated by an imaginary gap. The gap is located at the value $\Gamma=-2\gamma_\dc$ of the imaginary part $\Im \varepsilon(k)$ of the quasienergy spectrum. For each pair, the bands only cross at the invariant momenta $k=0,\pi$. Due to the Kramers degeneracy, this crossing persists when the red and blue sublattice are coupled for $J_\hc\ne 0$, see Fig.~\ref{Fig:2}(h). The quasienergy dispersion is symmetry-protected. As long as the symmetry relation~\eqref{TRS_t} is not broken and the imaginary gap does not close, the paired bands must wind around the quasienergy zone with opposite chirality.

For $J_\dc=\pi/2$, $J_\hc=\pi/2$, excitations starting on the open sites $\textcolor{red}{\circ}$, $\textcolor{blue}{\circ}$ move in a closed loop, leading to the dispersionless Kramers pair $\varepsilon_1^{I,II}(k)=-4\ii \gamma$. On the other hand, excitations that start on the filled sites $\textcolor{red}{\bullet}$, $\textcolor{blue}{\bullet}$ are transferred two unit cells to the right or left. Consequently, the two bands $\varepsilon_2^{I,II}(k)=\pm 2 k+\pi$ of the second Kramers pair wind twice around the quasienergy zone with two crossings at the non-invariant momenta $k=\pi/2$, $k=-\pi/2$. These crossings are not protected by Kramers degeneracy and thus are lifted for $J_\hc\ne \pi/2$, see Fig.~\ref{Fig:2}(i). As a result, the second Kramers pair no longer winds around the quasienergy zone. Note that the imaginary gap at $\Gamma=-2\gamma_\dc$ does not close during this process. Here, the winding around the quasienergy zone is not symmetry-protected. 

In general, the winding is only symmetry-protected  if the paired bands wind an odd number of times around the quasienergy zone with opposite chirality. The number of windings modulo $2$ is determined by the $\mathbb Z_2$-valued quantity
\begin{equation} \label{W1_TRS}
\begin{aligned}
W_\theta(\Gamma)
=\frac{1}{4\pi}\Bigg(&\sum_{\Gamma < \Im \varepsilon_m^I} \,  \big[ \Re  \varepsilon_m^I(k) \big]_{k=-\pi}^{k=\pi}  \\ -&\sum_{\Gamma < \Im \varepsilon_m^{II}}  \big[ \Re  \varepsilon_m^{II}(k) \big]_{k=-\pi}^{k=\pi} \Bigg)\mod 2\;,
\end{aligned}
\end{equation}
which includes all Kramers pairs above the imaginary gap $\ii \Gamma$. Eq.~\eqref{W1_TRS} generalizes the $\mathbb Z$-valued winding number $W(\Gamma)$ of Ref.~\onlinecite{hckendorf2019nonhermitian} to systems with  fermionic time-reversal symmetry.

\begin{figure}
\includegraphics[width=1\linewidth]{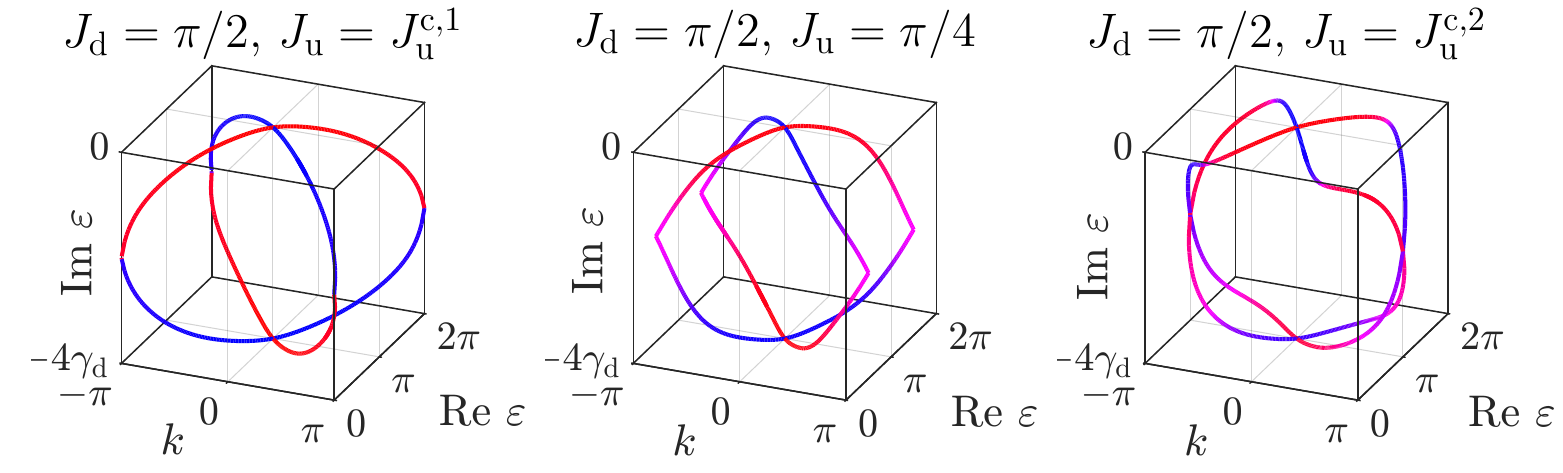}
\caption{ 
Quasienergies in the intermediate region that separates the non-trivial phase in Figs.~\ref{Fig:2}(f,h) from the trivial phase in Figs.~\ref{Fig:2}(g,i). We use $J_\dc=\pi/2$, $\gamma_\dc=0.5$ for all spectra so that $J_\hc^{\mathrm{c},1}\approx 0.43$, $J_\hc^{\mathrm{c},2}\approx 1.14$.  At the critical points $J_\hc^{\mathrm{c},1}$, $J_\hc^{\mathrm{c},2}$, the spectrum possesses an exceptional point at $k=\pm \pi$.
}
\label{Fig:3}
\end{figure}

For the trivial imaginary gap at $\Gamma=-\infty$, where all Kramers pairs in the quasienergy spectrum contribute, we have $W_\theta(-\infty)=0$. A single Kramers pair that winds an odd number of times around the quasienergy zone can not be generated by a time-reversal~symmetric Hamiltonian. This directly follows from the invertibility of the propagator. Going in inverse time direction from $t=T$ to $t=0$, the windings around the quasienergy zone can never disappear, but this contradicts the condition $\varepsilon(k)=0$  that the quasienergies satisfy at $t=0$. 
As a direct consequence, we have $W_\theta(\Gamma)=0$ for all $\Gamma$ in any  Hermitian Floquet chain, in agreement with the symmetry classification for Hermitian Floquet systems~\cite{PhysRevB.96.155118}.
Non-zero values of $W_\theta$ can only be achieved in a non-Hermitian Floquet chain where two Kramers pairs with non-trivial windings can be spectrally separated by an imaginary gap.

In the vicinity of $J_\dc=\pi/2$ and $J_\hc=0$ ($J_\hc=\pi/2$), we have $W(\Gamma)=1$ ($W(\Gamma)=0$) for the imaginary gap at $\Gamma=-2\gamma_\dc$. The two phases are separated by an intermediate region $J_\hc\in[J_\hc^{\mathrm c, 1},J_\hc^{\mathrm c,2}]$ in which the imaginary gap between the two Kramers pairs closes (see Fig.~\ref{Fig:3}). 
The two critical points $J_\hc^{\mathrm{c},1}\le J_\hc^{\mathrm{c},2}$ that mark the boundaries of the intermediate region are determined by the parameters $J_\dc$ and $\gamma_\dc$. For $J_\dc\ne 0$ and $e^{-\gamma_\dc}\to 0$, they coincide ($J_\hc^{\mathrm{c},1}=J_\hc^{\mathrm{c},2}=\pi/4$) and the intermediate region collapses to a single critical point. In the following, we will abbreviate the parameter values  $e^{-\gamma_\dc}\to 0$, $\lambda_\hc=\pi/4$ of this point  with $p$. They are the optimal parameters for the external control of topological transport.

\section{Controlling the direction of transport} 

\begin{figure}
\includegraphics[width=1\linewidth]{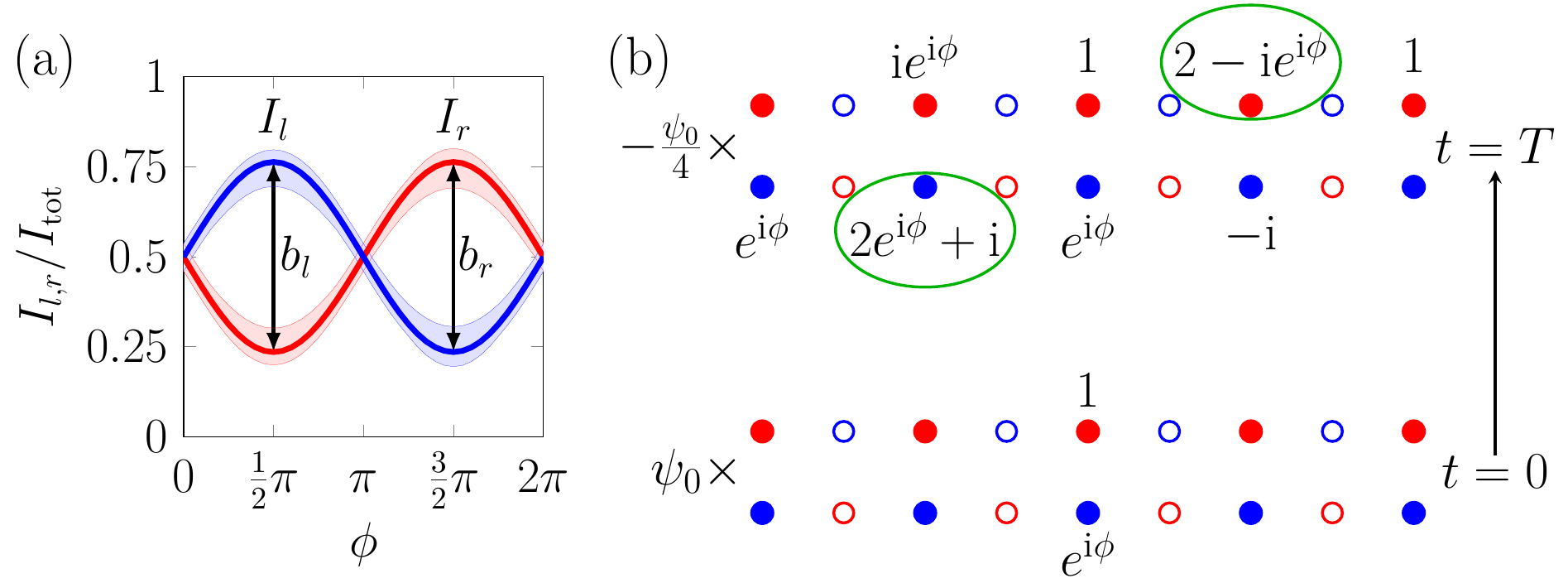}
\caption{(a) Calculated output intensities $I_l(\phi)$, $I_r(\phi)$ (solid curves) for the setup in Fig.\ref{Fig:1}(d) at $J_\dc=\pi/2$ and $p$. The initial state $\textcolor{blue}{|\Psi_0\rangle}$ is propagated over three cycles of the Floquet protocol. 
The shaded areas indicate the range in which the output intensities appear for $1000$ disorder configurations with $5\%$ uncertainty in the values of the coupling strengths.  (b) Distribution of the initial state  on the ladder after one Floquet cycle for zero disorder.
}
\label{Fig:4}
\end{figure}

The degree of control over the direction of transport in the setup of Fig.~\ref{Fig:1}(d) can be quantified with the peak intensity differences $b_l=\max (I_l(\phi)-I_r(\phi))/I_{\mathrm{tot}}$ and $b_r=\max (I_r(\phi)-I_l(\phi))/I_{\mathrm{tot}}$ of the two transport channels. The largest values of $b_{l,r}$ are realized at $p$, where the interference between excitations starting on adjacent filled sites $\textcolor{red}{\bullet}$, $\textcolor{blue}{\bullet}$ is maximized and the transport associated with the open sites $\textcolor{red}{\circ}$, $\textcolor{blue}{\circ}$ is fully suppressed. Here, we have $b_{l,r}\approx 0.5$, see Fig.~\ref{Fig:4}(a). The intensity $I_l(\phi)$ is approximately three times larger than $I_r(\phi)$ at $\phi=\pi/2$ and vice versa at $\phi=3\pi/2$. Note that the parameter $J_\dc$ does not influence the peak intensity differences $b_{l,r}$, but it determines the total output intensity $I_{\mathrm{tot}}$. In practice, $J_\dc$ should be set to $J_\dc=\pi/2$ to maximize $I_{\mathrm{tot}}$.

The position of the intensity minima and maxima at $\phi=\pi/2, 3\pi/2$ can be explained with the help of the Floquet propagator. At $J_\dc=\pi/2$ and $p$, the Floquet propagator $U(k)$ is a $4\times 4$ block diagonal matrix,
\begin{equation}
U(k)=\begin{pmatrix}
\label{propagator_full}
U_{\mathrm{fs}}(k) & 0 \\ 0 & 0 
\end{pmatrix}\; ,
\end{equation}
with the $2\times 2$ block
\begin{equation}
\label{propagator}
U_{\mathrm{fs}}(k)=-\frac{1}{4}\begin{pmatrix} 
1+2 e^{-\ii k}+e^{-2\ii k} & -2 \sin(k) \\ 2 \sin(k) & 1+2e^{\ii k}+e^{2\ii k}
\end{pmatrix}
\end{equation}
that acts on the filled sites and a $2\times 2$ block of zeroes that acts on the open sites. 
From Eq.~\eqref{propagator} it follows that the  initital state is distributed on the lattice at $t=T$ as shown in Fig.~\ref{Fig:4}(b). The interference that occurs on the two sites marked by green ellipses is controlled by the relative phase $\phi$. Constructive and destructive interference is maximized for $\phi=\pi/2, 3\pi/2$.

In a waveguide lattice, there is an uncertainty of approximately $5\%$ in the values of the coupling strengths $J_{\dc, \hc}$ for each pair of coupled waveguides~\cite{SzameitJPB}, which arises due to imperfections in the waveguide writing process. To test the robustness of the control of transport under these conditions, we replace the couplings, with strength $J_{\dc, \hc}$, by spatially and temporally disordered couplings, with strength $J_{\dc, \hc}^{\delta}=J_{\dc, \hc}+\delta$. Here, the disorder $\delta$ is drawn independently for each coupling from a uniform probability distribution in the intervall $[-0.95J_{\dc, \hc}, 1.05 J_{\dc, \hc}]$.  The resulting disorder in the Floquet ladder breaks time-reversal symmetry~\eqref{TRS_t} and the two transport channels are no longer symmetry-protected.  We calculate the output intensities $I_l(\phi)$, $I_r(\phi)$ for $1000$ disorder configurations. Averaged over all disorder configurations, we still have $b_{l,r}\approx 0.5$ with a standard deviation of approximately $8\%$. This demonstrates that the transport can be controlled reliably also in a real waveguide lattice with imperfections.

\begin{figure}
\includegraphics[width=1\linewidth]{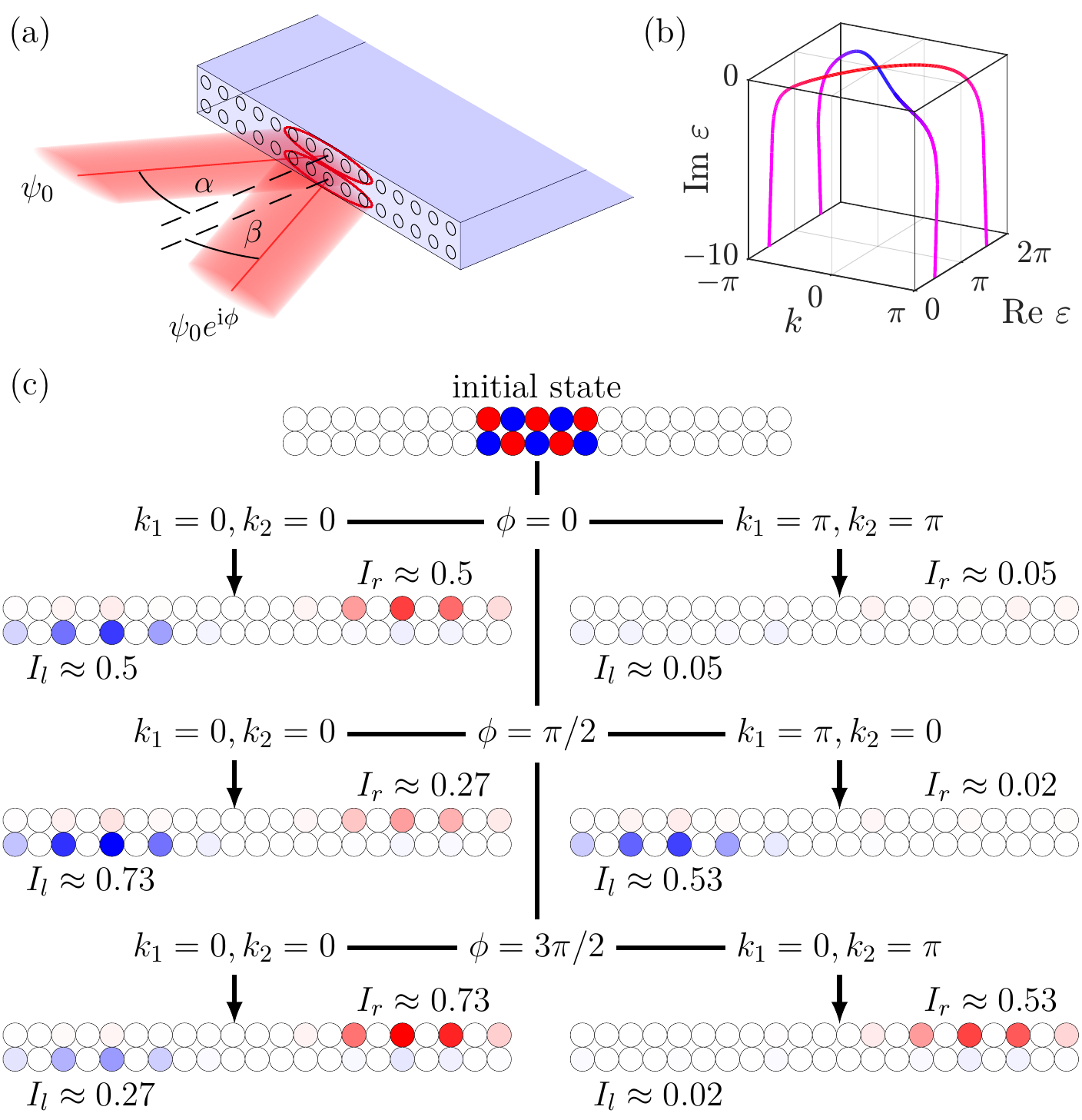}
\caption{(a) Enhanced setup for transport control. The component $\langle{\uparrow}|\textcolor{blue}{\Psi_0}\rangle$ [$\langle{\downarrow}|\textcolor{blue}{\Psi_0}\rangle$] of the inital state, that is located on the top row [bottom row] of the Floquet ladder, now has a specific momentum $k_1(\alpha)$ [$k_2(\beta)$]. The two momenta are determined by the angles of incidence $\alpha$ and $\beta$ of the two broad light beams. (b) Quasienergy dispersion~\eqref{eigenvalues} for the filled sites at $J_\dc=\pi/2$ and $p$. The imaginary part of the quasienergy diverges at $k=\pi$.  (c) Initial state $\textcolor{blue}{|\Psi_0\rangle}$ at $t=0$ and final states at $t=3T$ for different values of $\phi$, $k_1$, and $k_2$. The output intensities $I_{l,r}(\phi, k_1, k_2)$ are given in units of the total output intensity $I_{\mathrm{tot}}(0,0)$ at $k_{1,2}=0$. We use the parameters $J_\dc=\pi/2$ and $p$, with $5\%$ uncertainty in the values of the coupling strengths.
}
\label{Fig:5}
\end{figure}

The peak intensity differences $b_{l,r}$ can be further increased by exploiting the strong momentum-dependence of the losses in the ladder. The eigenvalues of the propagator~\eqref{propagator_full} are
\begin{equation}
\label{eigenvalues}
\begin{aligned}
e^{-\ii \varepsilon^{I,II}_{\mathrm{fs}}(k)}=&-\frac{1}{4}\bigg(2\cos(k)+2 \cos^2(k) \\
&\pm \ii \sin(k) \sqrt{8 \cos(k)+2\cos(2k)+10}\bigg)
\end{aligned}
\end{equation}
for the filled sites [see Fig.~\ref{Fig:5}(b)] and $e^{-\ii \varepsilon^{I,II}_{\mathrm{os}}(k)}=0$ for the open sites. Since the system is at the critical point, where the imaginary gap between the two Kramers pairs closes, we have $\Im \varepsilon_{\mathrm{fs}}^{I,II}(\pi)=-\infty$ at $k=\pi$. Any excitation with momentum $k=\pi$ gets fully suppressed during propagation. For $k=0$, on the other hand, we have $\Im \varepsilon^{I,II}_{\mathrm{fs}}(0)=0$, implying lossless propagation on the filled sites.

To generate states with specific momenta in a waveguide lattice, multiple waveguides have to be excited with a broad light beam under a specific angle of incidence~\cite{Maczewsky}. For the Floquet ladder, two separate momenta $k_1(\alpha)$, $k_2(\beta)$ can be introduced  into the initial state $\textcolor{blue}{\langle x|\Psi_0\rangle}=\psi_0(x) (e^{\ii k_1(\alpha)x}| {\uparrow}\rangle +e^{\ii k_2(\beta) x+\ii \phi} |{\downarrow}\rangle)$ by broadening the two input light beams along the horizontal axis and varying their angles of incidence $\alpha$, $\beta$ [see Fig.~\ref{Fig:5}(a)]. Note that the wavefunction $\psi_0(x)$ is now necessarily non-zero on multiple lattice sites.

The output intensities $I_l(\phi, k_1, k_2)$, $I_r(\phi, k_1, k_2)$ now depend on both the relative phase $\phi$ and the two momenta of the initial state.
Fig.~\ref{Fig:5}(c) provides a visual demonstration of the various transport manipulations that are possible when varying these three parameters. For $k_{1,2}=0$, only the components of the initial state that  start on the open sites will be suppressed during propagation and we recover the same $\phi$ dependence for the intensities $I_l(\phi,0,0)$, $I_r(\phi,0,0)$ as in Fig.~\ref{Fig:4}(a), with equal transport in both directions at $\phi=0$ and predominant transport to the left at $\phi=\pi/2$ (to the right at $\phi=3\pi/2$). As before, the total output intensity $I_{\mathrm{tot}}(k_1,k_2)$ does not depend on $\phi$. However, it depends on the two momenta.

For non-zero values of $k_1$ or $k_2$, the total output intensity  decreases. At $k_{1,2}=\pi$, where the components of the initial state that start on the filled sites are also suppressed, the total output intensity reaches its absolute minimum. Thus, any transport in the ladder is strongly suppressed. This is exemplarily shown for $\phi=0$ in Fig.~\ref{Fig:5}(c), but also applies to any other value of $\phi$.

If we set $k_1=\pi$ and $k_2=0$, the left-moving transport channel is suppressed significantly more than the right-moving transport channel and vice versa for  $k_1=0$ and $k_2=\pi$. This happens because the component $\langle{\uparrow}|\textcolor{blue}{\Psi_0}\rangle$  [$\langle{\downarrow}|\textcolor{blue}{\Psi_0}\rangle$] of the inital state contributes predominantly to the output intensity $I_l$ [$I_r$]. This effect and the unequal population of the two transport channels at $\phi=\pi/2, 3\pi/2$ can be combined to realize larger peak intensity differences $b_{l,r}$ than in Fig.~\ref{Fig:4}(a) at the cost of a lower value of $I_{\mathrm{tot}}$. 

Tab.~\ref{tab} summarizes the values of $b_{l,r}$ and $I_{\mathrm{tot}}$ for the two different setups. It also includes the case that the two momenta but not the relative phase are used for the control of transport. Which setup should be preferred depends on the specific application.

\begin{table}[]
\caption{Comparison between the three methods to externally control the direction of transport in the Floquet ladder. We use $J_\dc=\pi/2$ and $p$, with $5\%$ uncertainty in the values of the coupling strengths. The values in the table are calculated at $t=3T$ and averaged over $1000$ disorder configurations. We normalize the total output intensities such that $I_{\mathrm{tot}}=1$ in the first row of the table.}
\medskip
\begin{ruledtabular}
\begin{tabular}{c c c}
external parameters   & peak differences &  total intensity \\ \hline
relative phase $\phi$ & $b_{l,r}\approx0.5\phantom{0}$ &  $I_{\mathrm{tot}}=1$\\
angles of incidence $\alpha, \beta$ & $b_{l,r}\approx0.74$ & $I_{\mathrm{tot}}\approx0.55$\\
phase $\phi$ and angles $\alpha, \beta$  & $b_{l,r}\approx0.93$ &  $I_{\mathrm{tot}}\approx0.55$
\end{tabular}
\end{ruledtabular}
\label{tab}
\end{table}

\section{Conclusions}

To summarize, we have shown that the one-dimensional Floquet ladder introduced in this work provides a promising platform for the external control of the direction of topological transport. Instead of controlling transport by modification of the waveguide lattice, i.e. by re-engineering the Floquet ladder, the direction of transport is simply and reliably controlled by changing the relative phase between the two input light beams of the initial excitation. The phase sensitivity is a specific quality of the present setup, by which it is distinguished from simpler setups without time-reversal symmetry. There, counter-propagating transport channels have to remain uncoupled to preserve their topological properties. In effect, this means that the direction of transport is fixed entirely by the initial population of each of the separate transport channels and cannot resolve the relative phase  of the initial excitation.

Photonic waveguides provide a natural experimental setup for the implementation of our Floquet protocol and the realization of the transport control. The relative phase of the two input light beams can be controlled with a spatial light modulator as in, e.g., Ref.~\onlinecite{GreifRostock} and the momenta of the inital excitation can be controlled via the incidence angles of the two light beams. Our calculations demonstrate that the control of transport is unaffected by the disorder inherent to waveguide lattices. 

We have shown that the two distinct counter-propagating transport channels in the Floquet ladder are connected to a $\mathbb Z_2$ non-Hermitian Floquet topological phase, and are thus protected by time-reversal symmetry. Similar to the relationship between two-dimensional $\mathbb Z_2$ topological insulators and Chern insulators~\cite{HasanKane2010}, this $\mathbb Z_2$ phase is a time-reversal~symmetric generalization of the $\mathbb Z$ topological phase that was recently discovered in non-Hermitian Floquet chains with unidirectional transport channels~\cite{hckendorf2019nonhermitian, Fedorova2020}. In both phases an imaginary gap separates quasienergy bands which wind around the quasienergy zone. The separation through the imaginary gap contributes to the robustness of the observed effects, even beyond the time-reversal symmetric situation. In the $\mathbb Z$ phase, a large imaginary gap leads to quantized unidirectional transport~\cite{hckendorf2019nonhermitian}. For  the present system, it remains an open question whether the two counter-propagatingpropagating transport channels become quantized in the limit of a large imaginary gap.   

Interestingly, optimal conditions for the external control of transport are provided at the boundary of the $\mathbb Z_2$ phase, where the imaginary gap closes and the phase transition occurs. In a Hermitian Floquet system, this would indicate that transport is fragile, since a small perturbation can push the system into the topologically trivial phase. Here, in a non-Hermitian Floquet system, the transport remains robust when the system is pushed into the trivial phase because the imaginary part of the quasienergy is strongly momentum-dependent. Transport disappears only when this momentum-dependence flattens out~\cite{hckendorf2019nonhermitian}, which coincides with the closing of a point gap~\cite{PhysRevX.8.031079} in the quasienergy spectrum. 
This phenomenon highlights the intrinsic differences between topologically robust transport in Hermitian and non-Hermitian Floquet systems.

\section*{DATA AVAILABILITY}
The data that support the findings of this study are available from the corresponding author upon reasonable request.



%


\begin{thebibliography}{40}%
\makeatletter
\providecommand \@ifxundefined [1]{%
 \@ifx{#1\undefined}
}%
\providecommand \@ifnum [1]{%
 \ifnum #1\expandafter \@firstoftwo
 \else \expandafter \@secondoftwo
 \fi
}%
\providecommand \@ifx [1]{%
 \ifx #1\expandafter \@firstoftwo
 \else \expandafter \@secondoftwo
 \fi
}%
\providecommand \natexlab [1]{#1}%
\providecommand \enquote  [1]{``#1''}%
\providecommand \bibnamefont  [1]{#1}%
\providecommand \bibfnamefont [1]{#1}%
\providecommand \citenamefont [1]{#1}%
\providecommand \href@noop [0]{\@secondoftwo}%
\providecommand \href [0]{\begingroup \@sanitize@url \@href}%
\providecommand \@href[1]{\@@startlink{#1}\@@href}%
\providecommand \@@href[1]{\endgroup#1\@@endlink}%
\providecommand \@sanitize@url [0]{\catcode `\\12\catcode `\$12\catcode
  `\&12\catcode `\#12\catcode `\^12\catcode `\_12\catcode `\%12\relax}%
\providecommand \@@startlink[1]{}%
\providecommand \@@endlink[0]{}%
\providecommand \url  [0]{\begingroup\@sanitize@url \@url }%
\providecommand \@url [1]{\endgroup\@href {#1}{\urlprefix }}%
\providecommand \urlprefix  [0]{URL }%
\providecommand \Eprint [0]{\href }%
\providecommand \doibase [0]{http://dx.doi.org/}%
\providecommand \selectlanguage [0]{\@gobble}%
\providecommand \bibinfo  [0]{\@secondoftwo}%
\providecommand \bibfield  [0]{\@secondoftwo}%
\providecommand \translation [1]{[#1]}%
\providecommand \BibitemOpen [0]{}%
\providecommand \bibitemStop [0]{}%
\providecommand \bibitemNoStop [0]{.\EOS\space}%
\providecommand \EOS [0]{\spacefactor3000\relax}%
\providecommand \BibitemShut  [1]{\csname bibitem#1\endcsname}%
\let\auto@bib@innerbib\@empty
\bibitem [{\citenamefont {Hasan}\ and\ \citenamefont
  {Kane}(2010)}]{HasanKane2010}%
  \BibitemOpen
  \bibfield  {author} {\bibinfo {author} {\bibfnamefont {M.~Z.}\ \bibnamefont
  {Hasan}}\ and\ \bibinfo {author} {\bibfnamefont {C.~L.}\ \bibnamefont
  {Kane}},\ }\bibfield  {title} {\enquote {\bibinfo {title} {{Colloquium}:
  {Topological} insulators},}\ }\href {\doibase 10.1103/RevModPhys.82.3045}
  {\bibfield  {journal} {\bibinfo  {journal} {Rev. Mod. Phys.}\ }\textbf
  {\bibinfo {volume} {82}},\ \bibinfo {pages} {3045} (\bibinfo {year}
  {2010})}\BibitemShut {NoStop}%
\bibitem [{\citenamefont {Qi}\ and\ \citenamefont
  {Zhang}(2011)}]{RevModPhys.83.1057}%
  \BibitemOpen
  \bibfield  {author} {\bibinfo {author} {\bibfnamefont {X.-L.}\ \bibnamefont
  {Qi}}\ and\ \bibinfo {author} {\bibfnamefont {S.-C.}\ \bibnamefont {Zhang}},\
  }\bibfield  {title} {\enquote {\bibinfo {title} {Topological insulators and
  superconductors},}\ }\href {\doibase 10.1103/RevModPhys.83.1057} {\bibfield
  {journal} {\bibinfo  {journal} {Rev. Mod. Phys.}\ }\textbf {\bibinfo {volume}
  {83}},\ \bibinfo {pages} {1057} (\bibinfo {year} {2011})}\BibitemShut
  {NoStop}%
\bibitem [{\citenamefont {Chiu}\ \emph {et~al.}(2016)\citenamefont {Chiu},
  \citenamefont {Teo}, \citenamefont {Schnyder},\ and\ \citenamefont
  {Ryu}}]{RevModPhys.88.035005}%
  \BibitemOpen
  \bibfield  {author} {\bibinfo {author} {\bibfnamefont {C.-K.}\ \bibnamefont
  {Chiu}}, \bibinfo {author} {\bibfnamefont {J.~C.~Y.}\ \bibnamefont {Teo}},
  \bibinfo {author} {\bibfnamefont {A.~P.}\ \bibnamefont {Schnyder}}, \ and\
  \bibinfo {author} {\bibfnamefont {S.}~\bibnamefont {Ryu}},\ }\bibfield
  {title} {\enquote {\bibinfo {title} {Classification of topological quantum
  matter with symmetries},}\ }\href {\doibase 10.1103/RevModPhys.88.035005}
  {\bibfield  {journal} {\bibinfo  {journal} {Rev. Mod. Phys.}\ }\textbf
  {\bibinfo {volume} {88}},\ \bibinfo {pages} {035005} (\bibinfo {year}
  {2016})}\BibitemShut {NoStop}%
\bibitem [{\citenamefont {Klitzing}, \citenamefont {Dorda},\ and\ \citenamefont
  {Pepper}(1980)}]{Klitzing}%
  \BibitemOpen
  \bibfield  {author} {\bibinfo {author} {\bibfnamefont {K.~v.}\ \bibnamefont
  {Klitzing}}, \bibinfo {author} {\bibfnamefont {G.}~\bibnamefont {Dorda}}, \
  and\ \bibinfo {author} {\bibfnamefont {M.}~\bibnamefont {Pepper}},\
  }\bibfield  {title} {\enquote {\bibinfo {title} {New {M}ethod for
  {High-Accuracy Determination} of the {Fine-Structure Constant Based} on
  {Quantized Hall Resistance}},}\ }\href {\doibase 10.1103/PhysRevLett.45.494}
  {\bibfield  {journal} {\bibinfo  {journal} {Phys. Rev. Lett.}\ }\textbf
  {\bibinfo {volume} {45}},\ \bibinfo {pages} {494} (\bibinfo {year}
  {1980})}\BibitemShut {NoStop}%
\bibitem [{\citenamefont {K{\"o}nig}\ \emph {et~al.}(2007)\citenamefont
  {K{\"o}nig}, \citenamefont {Wiedmann}, \citenamefont {Br{\"u}ne},
  \citenamefont {Roth}, \citenamefont {Buhmann}, \citenamefont {Molenkamp},
  \citenamefont {Qi},\ and\ \citenamefont {Zhang}}]{Konig2007}%
  \BibitemOpen
  \bibfield  {author} {\bibinfo {author} {\bibfnamefont {M.}~\bibnamefont
  {K{\"o}nig}}, \bibinfo {author} {\bibfnamefont {S.}~\bibnamefont {Wiedmann}},
  \bibinfo {author} {\bibfnamefont {C.}~\bibnamefont {Br{\"u}ne}}, \bibinfo
  {author} {\bibfnamefont {A.}~\bibnamefont {Roth}}, \bibinfo {author}
  {\bibfnamefont {H.}~\bibnamefont {Buhmann}}, \bibinfo {author} {\bibfnamefont
  {L.~W.}\ \bibnamefont {Molenkamp}}, \bibinfo {author} {\bibfnamefont {X.-L.}\
  \bibnamefont {Qi}}, \ and\ \bibinfo {author} {\bibfnamefont {S.-C.}\
  \bibnamefont {Zhang}},\ }\bibfield  {title} {\enquote {\bibinfo {title}
  {{Quantum Spin Hall Insulator State} in {H}g{T}e {Quantum Wells}},}\ }\href
  {\doibase 10.1126/science.1148047} {\bibfield  {journal} {\bibinfo  {journal}
  {Science}\ }\textbf {\bibinfo {volume} {318}},\ \bibinfo {pages} {766}
  (\bibinfo {year} {2007})}\BibitemShut {NoStop}%
\bibitem [{\citenamefont {Hsieh}\ \emph {et~al.}(2008)\citenamefont {Hsieh},
  \citenamefont {Qian}, \citenamefont {Wray}, \citenamefont {Xia},
  \citenamefont {Hor}, \citenamefont {Cava},\ and\ \citenamefont
  {Hasan}}]{Hsieh}%
  \BibitemOpen
  \bibfield  {author} {\bibinfo {author} {\bibfnamefont {D.}~\bibnamefont
  {Hsieh}}, \bibinfo {author} {\bibfnamefont {D.}~\bibnamefont {Qian}},
  \bibinfo {author} {\bibfnamefont {L.}~\bibnamefont {Wray}}, \bibinfo {author}
  {\bibfnamefont {Y.}~\bibnamefont {Xia}}, \bibinfo {author} {\bibfnamefont
  {Y.~S.}\ \bibnamefont {Hor}}, \bibinfo {author} {\bibfnamefont {R.~J.}\
  \bibnamefont {Cava}}, \ and\ \bibinfo {author} {\bibfnamefont {M.~Z.}\
  \bibnamefont {Hasan}},\ }\bibfield  {title} {\enquote {\bibinfo {title} {A
  topological {D}irac insulator in a quantum spin {H}all phase},}\ }\href
  {http://dx.doi.org/10.1038/nature06843} {\bibfield  {journal} {\bibinfo
  {journal} {Nature}\ }\textbf {\bibinfo {volume} {452}},\ \bibinfo {pages}
  {970} (\bibinfo {year} {2008})}\BibitemShut {NoStop}%
\bibitem [{\citenamefont {Kane}\ and\ \citenamefont
  {Mele}(2005)}]{KaneMelePRL}%
  \BibitemOpen
  \bibfield  {author} {\bibinfo {author} {\bibfnamefont {C.~L.}\ \bibnamefont
  {Kane}}\ and\ \bibinfo {author} {\bibfnamefont {E.~J.}\ \bibnamefont
  {Mele}},\ }\bibfield  {title} {\enquote {\bibinfo {title} {${Z}_{2}$
  {Topological Order} and the {Quantum Spin Hall Effect}},}\ }\href {\doibase
  10.1103/PhysRevLett.95.146802} {\bibfield  {journal} {\bibinfo  {journal}
  {Phys. Rev. Lett.}\ }\textbf {\bibinfo {volume} {95}},\ \bibinfo {pages}
  {146802} (\bibinfo {year} {2005})}\BibitemShut {NoStop}%
\bibitem [{\citenamefont {Haldane}\ and\ \citenamefont
  {Raghu}(2008)}]{PhysRevLett.100.013904}%
  \BibitemOpen
  \bibfield  {author} {\bibinfo {author} {\bibfnamefont {F.~D.~M.}\
  \bibnamefont {Haldane}}\ and\ \bibinfo {author} {\bibfnamefont
  {S.}~\bibnamefont {Raghu}},\ }\bibfield  {title} {\enquote {\bibinfo {title}
  {Possible realization of directional optical waveguides in photonic crystals
  with broken time-reversal symmetry},}\ }\href {\doibase
  10.1103/PhysRevLett.100.013904} {\bibfield  {journal} {\bibinfo  {journal}
  {Phys. Rev. Lett.}\ }\textbf {\bibinfo {volume} {100}},\ \bibinfo {pages}
  {013904} (\bibinfo {year} {2008})}\BibitemShut {NoStop}%
\bibitem [{\citenamefont {Hafezi}\ \emph {et~al.}(2011)\citenamefont {Hafezi},
  \citenamefont {Demler}, \citenamefont {Lukin},\ and\ \citenamefont
  {Taylor}}]{Hafezi2011}%
  \BibitemOpen
  \bibfield  {author} {\bibinfo {author} {\bibfnamefont {M.}~\bibnamefont
  {Hafezi}}, \bibinfo {author} {\bibfnamefont {E.~A.}\ \bibnamefont {Demler}},
  \bibinfo {author} {\bibfnamefont {M.~D.}\ \bibnamefont {Lukin}}, \ and\
  \bibinfo {author} {\bibfnamefont {J.~M.}\ \bibnamefont {Taylor}},\ }\bibfield
   {title} {\enquote {\bibinfo {title} {Robust optical delay lines with
  topological protection},}\ }\href {https://doi.org/10.1038/nphys2063}
  {\bibfield  {journal} {\bibinfo  {journal} {Nat. Phys.}\ }\textbf {\bibinfo
  {volume} {7}},\ \bibinfo {pages} {907} (\bibinfo {year} {2011})} \BibitemShut {NoStop}%
\bibitem [{\citenamefont {Ozawa}\ \emph {et~al.}(2019)\citenamefont {Ozawa},
  \citenamefont {Price}, \citenamefont {Amo}, \citenamefont {Goldman},
  \citenamefont {Hafezi}, \citenamefont {Lu}, \citenamefont {Rechtsman},
  \citenamefont {Schuster}, \citenamefont {Simon}, \citenamefont {Zilberberg},\
  and\ \citenamefont {Carusotto}}]{RevModPhys.91.015006}%
  \BibitemOpen
  \bibfield  {author} {\bibinfo {author} {\bibfnamefont {T.}~\bibnamefont
  {Ozawa}}, \bibinfo {author} {\bibfnamefont {H.~M.}\ \bibnamefont {Price}},
  \bibinfo {author} {\bibfnamefont {A.}~\bibnamefont {Amo}}, \bibinfo {author}
  {\bibfnamefont {N.}~\bibnamefont {Goldman}}, \bibinfo {author} {\bibfnamefont
  {M.}~\bibnamefont {Hafezi}}, \bibinfo {author} {\bibfnamefont
  {L.}~\bibnamefont {Lu}}, \bibinfo {author} {\bibfnamefont {M.~C.}\
  \bibnamefont {Rechtsman}}, \bibinfo {author} {\bibfnamefont {D.}~\bibnamefont
  {Schuster}}, \bibinfo {author} {\bibfnamefont {J.}~\bibnamefont {Simon}},
  \bibinfo {author} {\bibfnamefont {O.}~\bibnamefont {Zilberberg}}, \ and\
  \bibinfo {author} {\bibfnamefont {I.}~\bibnamefont {Carusotto}},\ }\bibfield
  {title} {\enquote {\bibinfo {title} {Topological photonics},}\ }\href
  {\doibase 10.1103/RevModPhys.91.015006} {\bibfield  {journal} {\bibinfo
  {journal} {Rev. Mod. Phys.}\ }\textbf {\bibinfo {volume} {91}},\ \bibinfo
  {pages} {015006} (\bibinfo {year} {2019})}\BibitemShut {NoStop}%
\bibitem [{\citenamefont {Maczewsky}\ \emph {et~al.}(2020)\citenamefont
  {Maczewsky}, \citenamefont {H{\"o}ckendorf}, \citenamefont {Kremer},
  \citenamefont {Biesenthal}, \citenamefont {Heinrich}, \citenamefont
  {Alvermann}, \citenamefont {Fehske},\ and\ \citenamefont
  {Szameit}}]{GreifRostock}%
  \BibitemOpen
  \bibfield  {author} {\bibinfo {author} {\bibfnamefont {L.~J.}\ \bibnamefont
  {Maczewsky}}, \bibinfo {author} {\bibfnamefont {B.}~\bibnamefont
  {H{\"o}ckendorf}}, \bibinfo {author} {\bibfnamefont {M.}~\bibnamefont
  {Kremer}}, \bibinfo {author} {\bibfnamefont {T.}~\bibnamefont {Biesenthal}},
  \bibinfo {author} {\bibfnamefont {M.}~\bibnamefont {Heinrich}}, \bibinfo
  {author} {\bibfnamefont {A.}~\bibnamefont {Alvermann}}, \bibinfo {author}
  {\bibfnamefont {H.}~\bibnamefont {Fehske}}, \ and\ \bibinfo {author}
  {\bibfnamefont {A.}~\bibnamefont {Szameit}},\ }\bibfield  {title} {\enquote
  {\bibinfo {title} {Fermionic time-reversal symmetry in a photonic topological
  insulator},}\ }\href {\doibase 10.1038/s41563-020-0641-8} {\bibfield
  {journal} {\bibinfo  {journal} {Nat. Mater.}\ }\textbf {\bibinfo {volume}
  {19}},\ \bibinfo {pages} {855} (\bibinfo {year} {2020})}\BibitemShut
  {NoStop}%
\bibitem [{\citenamefont {Szameit}\ and\ \citenamefont
  {Nolte}(2010)}]{SzameitJPB}%
  \BibitemOpen
  \bibfield  {author} {\bibinfo {author} {\bibfnamefont {A.}~\bibnamefont
  {Szameit}}\ and\ \bibinfo {author} {\bibfnamefont {S.}~\bibnamefont
  {Nolte}},\ }\bibfield  {title} {\enquote {\bibinfo {title} {Discrete optics
  in femtosecond-laser-written photonic structures},}\ }\href
  {http://stacks.iop.org/0953-4075/43/i=16/a=163001} {\bibfield  {journal}
  {\bibinfo  {journal} {J. Phys. B}\ }\textbf {\bibinfo {volume} {43}},\
  \bibinfo {pages} {163001} (\bibinfo {year} {2010})}\BibitemShut {NoStop}%
\bibitem [{\citenamefont {Kitagawa}\ \emph {et~al.}(2010)\citenamefont
  {Kitagawa}, \citenamefont {Berg}, \citenamefont {Rudner},\ and\ \citenamefont
  {Demler}}]{KitagawaPRB}%
  \BibitemOpen
  \bibfield  {author} {\bibinfo {author} {\bibfnamefont {T.}~\bibnamefont
  {Kitagawa}}, \bibinfo {author} {\bibfnamefont {E.}~\bibnamefont {Berg}},
  \bibinfo {author} {\bibfnamefont {M.}~\bibnamefont {Rudner}}, \ and\ \bibinfo
  {author} {\bibfnamefont {E.}~\bibnamefont {Demler}},\ }\bibfield  {title}
  {\enquote {\bibinfo {title} {Topological characterization of periodically
  driven quantum systems},}\ }\href
  {http://link.aps.org/doi/10.1103/PhysRevB.82.235114} {\bibfield  {journal}
  {\bibinfo  {journal} {Phys. Rev. B}\ }\textbf {\bibinfo {volume} {82}},\
  \bibinfo {pages} {235114} (\bibinfo {year} {2010})}\BibitemShut {NoStop}%
\bibitem [{\citenamefont {Rudner}\ \emph {et~al.}(2013)\citenamefont {Rudner},
  \citenamefont {Lindner}, \citenamefont {Berg},\ and\ \citenamefont
  {Levin}}]{Rudner}%
  \BibitemOpen
  \bibfield  {author} {\bibinfo {author} {\bibfnamefont {M.~S.}\ \bibnamefont
  {Rudner}}, \bibinfo {author} {\bibfnamefont {N.~H.}\ \bibnamefont {Lindner}},
  \bibinfo {author} {\bibfnamefont {E.}~\bibnamefont {Berg}}, \ and\ \bibinfo
  {author} {\bibfnamefont {M.}~\bibnamefont {Levin}},\ }\bibfield  {title}
  {\enquote {\bibinfo {title} {{Anomalous Edge States} and the {Bulk-Edge
  Correspondence} for {Periodically Driven Two-Dimensional Systems}},}\ }\href
  {http://link.aps.org/doi/10.1103/PhysRevX.3.031005} {\bibfield  {journal}
  {\bibinfo  {journal} {Phys. Rev. X}\ }\textbf {\bibinfo {volume} {3}},\
  \bibinfo {pages} {031005} (\bibinfo {year} {2013})}\BibitemShut {NoStop}%
\bibitem [{\citenamefont {Maczewsky}\ \emph {et~al.}(2017)\citenamefont
  {Maczewsky}, \citenamefont {Zeuner}, \citenamefont {Nolte},\ and\
  \citenamefont {Szameit}}]{Maczewsky}%
  \BibitemOpen
  \bibfield  {author} {\bibinfo {author} {\bibfnamefont {L.~J.}\ \bibnamefont
  {Maczewsky}}, \bibinfo {author} {\bibfnamefont {J.~M.}\ \bibnamefont
  {Zeuner}}, \bibinfo {author} {\bibfnamefont {S.}~\bibnamefont {Nolte}}, \
  and\ \bibinfo {author} {\bibfnamefont {A.}~\bibnamefont {Szameit}},\
  }\bibfield  {title} {\enquote {\bibinfo {title} {Observation of photonic
  anomalous {F}loquet topological insulators},}\ }\href
  {http://dx.doi.org/10.1038/ncomms13756} {\bibfield  {journal} {\bibinfo
  {journal} {Nat. Comm.}\ }\textbf {\bibinfo {volume} {8}},\ \bibinfo {pages}
  {13756} (\bibinfo {year} {2017})}\BibitemShut {NoStop}%
\bibitem [{\citenamefont {Mukherjee}\ \emph {et~al.}(2017)\citenamefont
  {Mukherjee}, \citenamefont {Spracklen}, \citenamefont {Valiente},
  \citenamefont {Andersson}, \citenamefont {{\"O}hberg}, \citenamefont
  {Goldman},\ and\ \citenamefont {Thomson}}]{Mukherjee}%
  \BibitemOpen
  \bibfield  {author} {\bibinfo {author} {\bibfnamefont {S.}~\bibnamefont
  {Mukherjee}}, \bibinfo {author} {\bibfnamefont {A.}~\bibnamefont
  {Spracklen}}, \bibinfo {author} {\bibfnamefont {M.}~\bibnamefont {Valiente}},
  \bibinfo {author} {\bibfnamefont {E.}~\bibnamefont {Andersson}}, \bibinfo
  {author} {\bibfnamefont {P.}~\bibnamefont {{\"O}hberg}}, \bibinfo {author}
  {\bibfnamefont {N.}~\bibnamefont {Goldman}}, \ and\ \bibinfo {author}
  {\bibfnamefont {R.~R.}\ \bibnamefont {Thomson}},\ }\bibfield  {title}
  {\enquote {\bibinfo {title} {Experimental observation of anomalous
  topological edge modes in a slowly driven photonic lattice},}\ }\href
  {http://dx.doi.org/10.1038/ncomms13918} {\bibfield  {journal} {\bibinfo
  {journal} {Nat. Comm.}\ }\textbf {\bibinfo {volume} {8}},\ \bibinfo {pages}
  {13918} (\bibinfo {year} {2017})}\BibitemShut {NoStop}%
\bibitem [{\citenamefont {Rudner}\ and\ \citenamefont
  {Lindner}(2020)}]{Rudner2020}%
  \BibitemOpen
  \bibfield  {author} {\bibinfo {author} {\bibfnamefont {M.~S.}\ \bibnamefont
  {Rudner}}\ and\ \bibinfo {author} {\bibfnamefont {N.~H.}\ \bibnamefont
  {Lindner}},\ }\bibfield  {title} {\enquote {\bibinfo {title} {Band structure
  engineering and non-equilibrium dynamics in {F}loquet topological
  insulators},}\ }\href {\doibase 10.1038/s42254-020-0170-z} {\bibfield
  {journal} {\bibinfo  {journal} {Nat. Rev. Phys.}\ }\textbf {\bibinfo {volume}
  {2}},\ \bibinfo {pages} {229} (\bibinfo {year} {2020})}\BibitemShut {NoStop}%
\bibitem [{\citenamefont {Rechtsman}\ \emph {et~al.}(2013)\citenamefont
  {Rechtsman}, \citenamefont {Zeuner}, \citenamefont {Plotnik}, \citenamefont
  {Lumer}, \citenamefont {Podolsky}, \citenamefont {Dreisow}, \citenamefont
  {Nolte}, \citenamefont {Segev},\ and\ \citenamefont {Szameit}}]{Rechtsman}%
  \BibitemOpen
  \bibfield  {author} {\bibinfo {author} {\bibfnamefont {M.~C.}\ \bibnamefont
  {Rechtsman}}, \bibinfo {author} {\bibfnamefont {J.~M.}\ \bibnamefont
  {Zeuner}}, \bibinfo {author} {\bibfnamefont {Y.}~\bibnamefont {Plotnik}},
  \bibinfo {author} {\bibfnamefont {Y.}~\bibnamefont {Lumer}}, \bibinfo
  {author} {\bibfnamefont {D.}~\bibnamefont {Podolsky}}, \bibinfo {author}
  {\bibfnamefont {F.}~\bibnamefont {Dreisow}}, \bibinfo {author} {\bibfnamefont
  {S.}~\bibnamefont {Nolte}}, \bibinfo {author} {\bibfnamefont
  {M.}~\bibnamefont {Segev}}, \ and\ \bibinfo {author} {\bibfnamefont
  {A.}~\bibnamefont {Szameit}},\ }\bibfield  {title} {\enquote {\bibinfo
  {title} {Photonic {F}loquet topological insulators},}\ }\href
  {http://dx.doi.org/10.1038/nature12066} {\bibfield  {journal} {\bibinfo
  {journal} {Nature}\ }\textbf {\bibinfo {volume} {496}},\ \bibinfo {pages}
  {196} (\bibinfo {year} {2013})}\BibitemShut {NoStop}%
\bibitem [{\citenamefont {H\"ockendorf}, \citenamefont {Alvermann},\ and\
  \citenamefont {Fehske}(2019{\natexlab{a}})}]{HAF19}%
  \BibitemOpen
  \bibfield  {author} {\bibinfo {author} {\bibfnamefont {B.}~\bibnamefont
  {H\"ockendorf}}, \bibinfo {author} {\bibfnamefont {A.}~\bibnamefont
  {Alvermann}}, \ and\ \bibinfo {author} {\bibfnamefont {H.}~\bibnamefont
  {Fehske}},\ }\bibfield  {title} {\enquote {\bibinfo {title} {Universal
  driving protocol for symmetry-protected {F}loquet topological phases},}\
  }\href {\doibase 10.1103/PhysRevB.99.245102} {\bibfield  {journal} {\bibinfo
  {journal} {Phys. Rev. B}\ }\textbf {\bibinfo {volume} {99}},\ \bibinfo
  {pages} {245102} (\bibinfo {year} {2019}{\natexlab{a}})}\BibitemShut
  {NoStop}%
\bibitem [{\citenamefont {Weimann}\ \emph {et~al.}(2016)\citenamefont
  {Weimann}, \citenamefont {Kremer}, \citenamefont {Plotnik}, \citenamefont
  {Lumer}, \citenamefont {Nolte}, \citenamefont {Makris}, \citenamefont
  {Segev}, \citenamefont {Rechtsman},\ and\ \citenamefont
  {Szameit}}]{Weimann2016}%
  \BibitemOpen
  \bibfield  {author} {\bibinfo {author} {\bibfnamefont {S.}~\bibnamefont
  {Weimann}}, \bibinfo {author} {\bibfnamefont {M.}~\bibnamefont {Kremer}},
  \bibinfo {author} {\bibfnamefont {Y.}~\bibnamefont {Plotnik}}, \bibinfo
  {author} {\bibfnamefont {Y.}~\bibnamefont {Lumer}}, \bibinfo {author}
  {\bibfnamefont {S.}~\bibnamefont {Nolte}}, \bibinfo {author} {\bibfnamefont
  {K.~G.}\ \bibnamefont {Makris}}, \bibinfo {author} {\bibfnamefont
  {M.}~\bibnamefont {Segev}}, \bibinfo {author} {\bibfnamefont {M.~C.}\
  \bibnamefont {Rechtsman}}, \ and\ \bibinfo {author} {\bibfnamefont
  {A.}~\bibnamefont {Szameit}},\ }\bibfield  {title} {\enquote {\bibinfo
  {title} {Topologically protected bound states in photonic
  parity-time-symmetric crystals},}\ }\href {https://doi.org/10.1038/nmat4811}
  {\bibfield  {journal} {\bibinfo  {journal} {Nat. Mater.}\ }\textbf {\bibinfo
  {volume} {16}},\ \bibinfo {pages} {433} (\bibinfo {year} {2016})}\BibitemShut
  {NoStop}%
\bibitem [{\citenamefont {Weidemann}\ \emph {et~al.}(2020)\citenamefont
  {Weidemann}, \citenamefont {Kremer}, \citenamefont {Helbig}, \citenamefont
  {Hofmann}, \citenamefont {Stegmaier}, \citenamefont {Greiter}, \citenamefont
  {Thomale},\ and\ \citenamefont {Szameit}}]{Weidemann311}%
  \BibitemOpen
  \bibfield  {author} {\bibinfo {author} {\bibfnamefont {S.}~\bibnamefont
  {Weidemann}}, \bibinfo {author} {\bibfnamefont {M.}~\bibnamefont {Kremer}},
  \bibinfo {author} {\bibfnamefont {T.}~\bibnamefont {Helbig}}, \bibinfo
  {author} {\bibfnamefont {T.}~\bibnamefont {Hofmann}}, \bibinfo {author}
  {\bibfnamefont {A.}~\bibnamefont {Stegmaier}}, \bibinfo {author}
  {\bibfnamefont {M.}~\bibnamefont {Greiter}}, \bibinfo {author} {\bibfnamefont
  {R.}~\bibnamefont {Thomale}}, \ and\ \bibinfo {author} {\bibfnamefont
  {A.}~\bibnamefont {Szameit}},\ }\bibfield  {title} {\enquote {\bibinfo
  {title} {Topological funneling of light},}\ }\href {\doibase
  10.1126/science.aaz8727} {\bibfield  {journal} {\bibinfo  {journal}
  {Science}\ }\textbf {\bibinfo {volume} {368}},\ \bibinfo {pages} {311}
  (\bibinfo {year} {2020})}\BibitemShut {NoStop}%
\bibitem [{\citenamefont {Mukherjee}\ and\ \citenamefont
  {Rechtsman}(2020{\natexlab{a}})}]{Mukherjee856}%
  \BibitemOpen
  \bibfield  {author} {\bibinfo {author} {\bibfnamefont {S.}~\bibnamefont
  {Mukherjee}}\ and\ \bibinfo {author} {\bibfnamefont {M.~C.}\ \bibnamefont
  {Rechtsman}},\ }\bibfield  {title} {\enquote {\bibinfo {title} {Observation
  of {F}loquet solitons in a topological bandgap},}\ }\href {\doibase
  10.1126/science.aba8725} {\bibfield  {journal} {\bibinfo  {journal}
  {Science}\ }\textbf {\bibinfo {volume} {368}},\ \bibinfo {pages} {856}
  (\bibinfo {year} {2020}{\natexlab{a}})}\BibitemShut {NoStop}%
\bibitem [{\citenamefont {Mukherjee}\ and\ \citenamefont
  {Rechtsman}(2020{\natexlab{b}})}]{mukherjee2020observation}%
  \BibitemOpen
  \bibfield  {author} {\bibinfo {author} {\bibfnamefont {S.}~\bibnamefont
  {Mukherjee}}\ and\ \bibinfo {author} {\bibfnamefont {M.~C.}\ \bibnamefont
  {Rechtsman}},\ }\bibfield  {title} {\enquote {\bibinfo {title} {Observation
  of unidirectional soliton-like edge states in nonlinear {F}loquet topological
  insulators},}\ }\href {https://arxiv.org/abs/2010.11359} {\bibfield
  {journal} {\bibinfo  {journal} {arXiv:2010.11359}\ } (\bibinfo {year}
  {2020}{\natexlab{b}})}\BibitemShut {NoStop}%
\bibitem [{\citenamefont {Gong}\ \emph {et~al.}(2018)\citenamefont {Gong},
  \citenamefont {Ashida}, \citenamefont {Kawabata}, \citenamefont {Takasan},
  \citenamefont {Higashikawa},\ and\ \citenamefont {Ueda}}]{PhysRevX.8.031079}%
  \BibitemOpen
  \bibfield  {author} {\bibinfo {author} {\bibfnamefont {Z.}~\bibnamefont
  {Gong}}, \bibinfo {author} {\bibfnamefont {Y.}~\bibnamefont {Ashida}},
  \bibinfo {author} {\bibfnamefont {K.}~\bibnamefont {Kawabata}}, \bibinfo
  {author} {\bibfnamefont {K.}~\bibnamefont {Takasan}}, \bibinfo {author}
  {\bibfnamefont {S.}~\bibnamefont {Higashikawa}}, \ and\ \bibinfo {author}
  {\bibfnamefont {M.}~\bibnamefont {Ueda}},\ }\bibfield  {title} {\enquote
  {\bibinfo {title} {Topological Phases of Non-Hermitian systems},}\ }\href
  {\doibase 10.1103/PhysRevX.8.031079} {\bibfield  {journal} {\bibinfo
  {journal} {Phys. Rev. X}\ }\textbf {\bibinfo {volume} {8}},\ \bibinfo {pages}
  {031079} (\bibinfo {year} {2018})}\BibitemShut {NoStop}%
\bibitem [{\citenamefont {Kawabata}\ \emph {et~al.}(2019)\citenamefont
  {Kawabata}, \citenamefont {Shiozaki}, \citenamefont {Ueda},\ and\
  \citenamefont {Sato}}]{PhysRevX.9.041015}%
  \BibitemOpen
  \bibfield  {author} {\bibinfo {author} {\bibfnamefont {K.}~\bibnamefont
  {Kawabata}}, \bibinfo {author} {\bibfnamefont {K.}~\bibnamefont {Shiozaki}},
  \bibinfo {author} {\bibfnamefont {M.}~\bibnamefont {Ueda}}, \ and\ \bibinfo
  {author} {\bibfnamefont {M.}~\bibnamefont {Sato}},\ }\bibfield  {title}
  {\enquote {\bibinfo {title} {Symmetry and {T}opology in {N}on-{H}ermitian
  {P}hysics},}\ }\href {\doibase 10.1103/PhysRevX.9.041015} {\bibfield
  {journal} {\bibinfo  {journal} {Phys. Rev. X}\ }\textbf {\bibinfo {volume}
  {9}},\ \bibinfo {pages} {041015} (\bibinfo {year} {2019})}\BibitemShut
  {NoStop}%
\bibitem [{\citenamefont {Zhou}\ and\ \citenamefont
  {Lee}(2019)}]{PhysRevB.99.235112}%
  \BibitemOpen
  \bibfield  {author} {\bibinfo {author} {\bibfnamefont {H.}~\bibnamefont
  {Zhou}}\ and\ \bibinfo {author} {\bibfnamefont {J.~Y.}\ \bibnamefont {Lee}},\
  }\bibfield  {title} {\enquote {\bibinfo {title} {Periodic table for
  topological bands with non-{H}ermitian symmetries},}\ }\href {\doibase
  10.1103/PhysRevB.99.235112} {\bibfield  {journal} {\bibinfo  {journal} {Phys.
  Rev. B}\ }\textbf {\bibinfo {volume} {99}},\ \bibinfo {pages} {235112}
  (\bibinfo {year} {2019})}\BibitemShut {NoStop}%
\bibitem [{\citenamefont {Lee}\ \emph {et~al.}(2019)\citenamefont {Lee},
  \citenamefont {Ahn}, \citenamefont {Zhou},\ and\ \citenamefont
  {Vishwanath}}]{PhysRevLett.123.206404}%
  \BibitemOpen
  \bibfield  {author} {\bibinfo {author} {\bibfnamefont {J.~Y.}\ \bibnamefont
  {Lee}}, \bibinfo {author} {\bibfnamefont {J.}~\bibnamefont {Ahn}}, \bibinfo
  {author} {\bibfnamefont {H.}~\bibnamefont {Zhou}}, \ and\ \bibinfo {author}
  {\bibfnamefont {A.}~\bibnamefont {Vishwanath}},\ }\bibfield  {title}
  {\enquote {\bibinfo {title} {Topological {C}orrespondence between {H}ermitian
  and {N}on-{H}ermitian {S}ystems: {A}nomalous {D}ynamics},}\ }\href {\doibase
  10.1103/PhysRevLett.123.206404} {\bibfield  {journal} {\bibinfo  {journal}
  {Phys. Rev. Lett.}\ }\textbf {\bibinfo {volume} {123}},\ \bibinfo {pages}
  {206404} (\bibinfo {year} {2019})}\BibitemShut {NoStop}%
\bibitem [{\citenamefont {Bergholtz}, \citenamefont {Budich},\ and\
  \citenamefont {Kunst}(2020)}]{bergholtz2020exceptional}%
  \BibitemOpen
  \bibfield  {author} {\bibinfo {author} {\bibfnamefont {E.~J.}\ \bibnamefont
  {Bergholtz}}, \bibinfo {author} {\bibfnamefont {J.~C.}\ \bibnamefont
  {Budich}}, \ and\ \bibinfo {author} {\bibfnamefont {F.~K.}\ \bibnamefont
  {Kunst}},\ }\bibfield  {title} {\enquote {\bibinfo {title} {Exceptional
  {T}opology of {N}on-{H}ermitian {S}ystems},}\ }\href
  {https://arxiv.org/abs/1912.10048} {\bibfield  {journal} {\bibinfo  {journal}
  {arXiv:1912.10048}\ } (\bibinfo {year} {2020})}\BibitemShut {NoStop}%
\bibitem [{\citenamefont {Zhou}\ and\ \citenamefont
  {Gong}(2018)}]{PhysRevB.98.205417}%
  \BibitemOpen
  \bibfield  {author} {\bibinfo {author} {\bibfnamefont {L.}~\bibnamefont
  {Zhou}}\ and\ \bibinfo {author} {\bibfnamefont {J.}~\bibnamefont {Gong}},\
  }\bibfield  {title} {\enquote {\bibinfo {title} {Non-{H}ermitian {F}loquet
  topological phases with arbitrarily many real-quasienergy edge states},}\
  }\href {\doibase 10.1103/PhysRevB.98.205417} {\bibfield  {journal} {\bibinfo
  {journal} {Phys. Rev. B}\ }\textbf {\bibinfo {volume} {98}},\ \bibinfo
  {pages} {205417} (\bibinfo {year} {2018})}\BibitemShut {NoStop}%
\bibitem [{\citenamefont {H\"ockendorf}, \citenamefont {Alvermann},\ and\
  \citenamefont {Fehske}(2019{\natexlab{b}})}]{PhysRevLett.123.190403}%
  \BibitemOpen
  \bibfield  {author} {\bibinfo {author} {\bibfnamefont {B.}~\bibnamefont
  {H\"ockendorf}}, \bibinfo {author} {\bibfnamefont {A.}~\bibnamefont
  {Alvermann}}, \ and\ \bibinfo {author} {\bibfnamefont {H.}~\bibnamefont
  {Fehske}},\ }\bibfield  {title} {\enquote {\bibinfo {title} {Non-{H}ermitian
  {B}oundary {S}tate {E}ngineering in {A}nomalous {F}loquet {T}opological
  {I}nsulators},}\ }\href {\doibase 10.1103/PhysRevLett.123.190403} {\bibfield
  {journal} {\bibinfo  {journal} {Phys. Rev. Lett.}\ }\textbf {\bibinfo
  {volume} {123}},\ \bibinfo {pages} {190403} (\bibinfo {year}
  {2019}{\natexlab{b}})}\BibitemShut {NoStop}%
\bibitem [{\citenamefont {Zhou}\ and\ \citenamefont
  {Pan}(2019)}]{PhysRevA.100.053608}%
  \BibitemOpen
  \bibfield  {author} {\bibinfo {author} {\bibfnamefont {L.}~\bibnamefont
  {Zhou}}\ and\ \bibinfo {author} {\bibfnamefont {J.}~\bibnamefont {Pan}},\
  }\bibfield  {title} {\enquote {\bibinfo {title} {Non-{H}ermitian {F}loquet
  topological phases in the double-kicked rotor},}\ }\href {\doibase
  10.1103/PhysRevA.100.053608} {\bibfield  {journal} {\bibinfo  {journal}
  {Phys. Rev. A}\ }\textbf {\bibinfo {volume} {100}},\ \bibinfo {pages}
  {053608} (\bibinfo {year} {2019})}\BibitemShut {NoStop}%
\bibitem [{\citenamefont {Li}\ \emph {et~al.}(2019)\citenamefont {Li},
  \citenamefont {Ni}, \citenamefont {Weiner}, \citenamefont {Al\`u},\ and\
  \citenamefont {Khanikaev}}]{PhysRevB.100.045423}%
  \BibitemOpen
  \bibfield  {author} {\bibinfo {author} {\bibfnamefont {M.}~\bibnamefont
  {Li}}, \bibinfo {author} {\bibfnamefont {X.}~\bibnamefont {Ni}}, \bibinfo
  {author} {\bibfnamefont {M.}~\bibnamefont {Weiner}}, \bibinfo {author}
  {\bibfnamefont {A.}~\bibnamefont {Al\`u}}, \ and\ \bibinfo {author}
  {\bibfnamefont {A.~B.}\ \bibnamefont {Khanikaev}},\ }\bibfield  {title}
  {\enquote {\bibinfo {title} {Topological phases and nonreciprocal edge states
  in non-{H}ermitian {F}loquet insulators},}\ }\href {\doibase
  10.1103/PhysRevB.100.045423} {\bibfield  {journal} {\bibinfo  {journal}
  {Phys. Rev. B}\ }\textbf {\bibinfo {volume} {100}},\ \bibinfo {pages}
  {045423} (\bibinfo {year} {2019})}\BibitemShut {NoStop}%
\bibitem [{\citenamefont {H\"ockendorf}, \citenamefont {Alvermann},\ and\
  \citenamefont {Fehske}(2020)}]{hckendorf2019nonhermitian}%
  \BibitemOpen
  \bibfield  {author} {\bibinfo {author} {\bibfnamefont {B.}~\bibnamefont
  {H\"ockendorf}}, \bibinfo {author} {\bibfnamefont {A.}~\bibnamefont
  {Alvermann}}, \ and\ \bibinfo {author} {\bibfnamefont {H.}~\bibnamefont
  {Fehske}},\ }\bibfield  {title} {\enquote {\bibinfo {title} {Topological
  origin of quantized transport in non-{H}ermitian {F}loquet chains},}\ }\href
  {\doibase 10.1103/PhysRevResearch.2.023235} {\bibfield  {journal} {\bibinfo
  {journal} {Phys. Rev. Research}\ }\textbf {\bibinfo {volume} {2}},\ \bibinfo
  {pages} {023235} (\bibinfo {year} {2020})}\BibitemShut {NoStop}%
\bibitem [{\citenamefont {Fedorova}\ \emph {et~al.}(2020)\citenamefont
  {Fedorova}, \citenamefont {Qiu}, \citenamefont {Linden},\ and\ \citenamefont
  {Kroha}}]{Fedorova2020}%
  \BibitemOpen
  \bibfield  {author} {\bibinfo {author} {\bibfnamefont {Z.}~\bibnamefont
  {Fedorova}}, \bibinfo {author} {\bibfnamefont {H.}~\bibnamefont {Qiu}},
  \bibinfo {author} {\bibfnamefont {S.}~\bibnamefont {Linden}}, \ and\ \bibinfo
  {author} {\bibfnamefont {J.}~\bibnamefont {Kroha}},\ }\bibfield  {title}
  {\enquote {\bibinfo {title} {Observation of topological transport
  quantization by dissipation in fast {T}houless pumps},}\ }\href {\doibase
  10.1038/s41467-020-17510-z} {\bibfield  {journal} {\bibinfo  {journal} {Nat.
  Comm.}\ }\textbf {\bibinfo {volume} {11}},\ \bibinfo {pages} {3758} (\bibinfo
  {year} {2020})}\BibitemShut {NoStop}%
\bibitem [{\citenamefont {Zhang}\ and\ \citenamefont
  {Gong}(2020)}]{PhysRevB.101.045415}%
  \BibitemOpen
  \bibfield  {author} {\bibinfo {author} {\bibfnamefont {X.}~\bibnamefont
  {Zhang}}\ and\ \bibinfo {author} {\bibfnamefont {J.}~\bibnamefont {Gong}},\
  }\bibfield  {title} {\enquote {\bibinfo {title} {Non-{H}ermitian {F}loquet
  topological phases: Exceptional points, coalescent edge modes, and the skin
  effect},}\ }\href {\doibase 10.1103/PhysRevB.101.045415} {\bibfield
  {journal} {\bibinfo  {journal} {Phys. Rev. B}\ }\textbf {\bibinfo {volume}
  {101}},\ \bibinfo {pages} {045415} (\bibinfo {year} {2020})}\BibitemShut
  {NoStop}%
\bibitem [{\citenamefont {Wu}\ and\ \citenamefont
  {An}(2020)}]{PhysRevB.102.041119}%
  \BibitemOpen
  \bibfield  {author} {\bibinfo {author} {\bibfnamefont {H.}~\bibnamefont
  {Wu}}\ and\ \bibinfo {author} {\bibfnamefont {J.-H.}\ \bibnamefont {An}},\
  }\bibfield  {title} {\enquote {\bibinfo {title} {Floquet topological phases
  of non-{H}ermitian systems},}\ }\href {\doibase 10.1103/PhysRevB.102.041119}
  {\bibfield  {journal} {\bibinfo  {journal} {Phys. Rev. B}\ }\textbf {\bibinfo
  {volume} {102}},\ \bibinfo {pages} {041119} (\bibinfo {year}
  {2020})}\BibitemShut {NoStop}%
\bibitem [{\citenamefont {Zhou}(2020)}]{PhysRevB.101.014306}%
  \BibitemOpen
  \bibfield  {author} {\bibinfo {author} {\bibfnamefont {L.}~\bibnamefont
  {Zhou}},\ }\bibfield  {title} {\enquote {\bibinfo {title} {Non-{H}ermitian
  {F}loquet topological superconductors with multiple {M}ajorana edge modes},}\
  }\href {\doibase 10.1103/PhysRevB.101.014306} {\bibfield  {journal} {\bibinfo
   {journal} {Phys. Rev. B}\ }\textbf {\bibinfo {volume} {101}},\ \bibinfo
  {pages} {014306} (\bibinfo {year} {2020})}\BibitemShut {NoStop}%
\bibitem [{\citenamefont {Höckendorf}, \citenamefont {Alvermann},\ and\
  \citenamefont {Fehske}(2020)}]{H_ckendorf_2020}%
  \BibitemOpen
  \bibfield  {author} {\bibinfo {author} {\bibfnamefont {B.}~\bibnamefont
  {Höckendorf}}, \bibinfo {author} {\bibfnamefont {A.}~\bibnamefont
  {Alvermann}}, \ and\ \bibinfo {author} {\bibfnamefont {H.}~\bibnamefont
  {Fehske}},\ }\bibfield  {title} {\enquote {\bibinfo {title} {Cutting off the
  non-{H}ermitian boundary from an anomalous {F}loquet topological
  insulator},}\ }\href {\doibase 10.1209/0295-5075/131/30007} {\bibfield
  {journal} {\bibinfo  {journal} {Europhys. Lett.}\ }\textbf {\bibinfo {volume}
  {131}},\ \bibinfo {pages} {30007} (\bibinfo {year} {2020})}\BibitemShut
  {NoStop}%
\bibitem [{\citenamefont {Ornigotti}\ and\ \citenamefont
  {Szameit}(2014)}]{Ornigotti_2014}%
  \BibitemOpen
  \bibfield  {author} {\bibinfo {author} {\bibfnamefont {M.}~\bibnamefont
  {Ornigotti}}\ and\ \bibinfo {author} {\bibfnamefont {A.}~\bibnamefont
  {Szameit}},\ }\bibfield  {title} {\enquote {\bibinfo {title} {Quasi
  {P}{T}-symmetry in passive photonic lattices},}\ }\href {\doibase
  10.1088/2040-8978/16/6/065501} {\bibfield  {journal} {\bibinfo  {journal} {J.
  Opt.}\ }\textbf {\bibinfo {volume} {16}},\ \bibinfo {pages} {065501}
  (\bibinfo {year} {2014})}\BibitemShut {NoStop}%
\bibitem [{\citenamefont {Roy}\ and\ \citenamefont
  {Harper}(2017)}]{PhysRevB.96.155118}%
  \BibitemOpen
  \bibfield  {author} {\bibinfo {author} {\bibfnamefont {R.}~\bibnamefont
  {Roy}}\ and\ \bibinfo {author} {\bibfnamefont {F.}~\bibnamefont {Harper}},\
  }\bibfield  {title} {\enquote {\bibinfo {title} {Periodic table for {F}loquet
  topological insulators},}\ }\href {\doibase 10.1103/PhysRevB.96.155118}
  {\bibfield  {journal} {\bibinfo  {journal} {Phys. Rev. B}\ }\textbf {\bibinfo
  {volume} {96}},\ \bibinfo {pages} {155118} (\bibinfo {year}
  {2017})}\BibitemShut {NoStop}%
\end{thebibliography}
\end{document}